\documentclass[fleqn,usenatbib]{mnras}

\usepackage{newtxtext,newtxmath}

\usepackage[T1]{fontenc}

\usepackage{graphicx}	
\usepackage{amsmath}	
\usepackage{xcolor}
\usepackage{caption}
\usepackage{subcaption}

\title[General relativistic moving-mesh hydrodynamics]{General relativistic moving-mesh hydrodynamics simulations with \textsc{Arepo} and applications to neutron star mergers}

\author[Lioutas et al.]{Georgios Lioutas$^{1,2}$\thanks{E-mail: g.lioutas@gsi.de},
 Andreas Bauswein$^{1,3}$,
 Theodoros Soultanis$^{1,4,5}$,
 R\"udiger Pakmor$^{6}$,
 Volker Springel$^{6}$,
 \newauthor
 Friedrich K. R\"opke$^{4,7}$
\\
$^{1}$GSI Helmholtzzentrum f\"ur Schwerionenforschung, Planckstra{\ss}e 1, 64291 Darmstadt, Germany\\
$^{2}$Department of Physics and Astronomy, Ruprecht-Karls-Universit{\"a}t Heidelberg, Im Neuenheimer Feld 226, 69120 Heidelberg, Germany\\
$^{3}$Helmholtz Research Academy Hesse for FAIR (HFHF), GSI Helmholtz Center for Heavy Ion Research, Campus Darmstadt, Germany\\
$^{4}$Heidelberger Institut f\"ur Theoretische Studien, Schloss-Wolfsbrunnenweg 35, 69118, Heidelberg, Germany\\
$^{5}$Max-Planck-Institut f\"ur Astronomie, K\"onigstuhl 17, 69117 Heidelberg, Germany\\
$^{6}$Max-Planck-Institut f\"{u}r Astrophysik, Karl-Schwarzschild-Str. 1, D-85748, Garching, Germany\\
$^{7}$Zentrum f{\"u}r Astronomie der Universit{\"a}t Heidelberg, Institut f{\"u}r Theoretische Astrophysik, Philosophenweg 12, D-69120 Heidelberg, Germany
}

\date{Accepted XXX. Received YYY; in original form ZZZ}

\pubyear{2022}

\begin{document}
\label{firstpage}
\pagerange{\pageref{firstpage}--\pageref{lastpage}}
\maketitle

\begin{abstract}
We implement general relativistic hydrodynamics in the moving-mesh code \textsc{Arepo}. We also couple a solver for the Einstein field equations employing the conformal flatness approximation. The implementation is validated by evolving isolated static neutron stars using a fixed metric or a dynamical spacetime. In both tests the frequencies of the radial oscillation mode match those of independent calculations. We run the first moving-mesh simulation of a neutron star merger. The simulation includes a scheme to adaptively refine or derefine cells and thereby adjusting the local resolution dynamically. The general dynamics are in agreement with independent smoothed particle hydrodynamics and static-mesh simulations of neutron star mergers. Coarsely comparing, we find that dynamical features like the post-merger double-core structure or the quasi-radial oscillation mode persist on longer time scales, possibly reflecting a low numerical diffusivity of our method. Similarly, the post-merger gravitational wave emission shows the same features as observed in simulations with other codes. In particular, the main frequency of the post-merger phase is found to be in good agreement with independent results for the same binary system, while, in comparison, the amplitude of the post-merger gravitational wave signal falls off slower, i.e. the post-merger oscillations are less damped. The successful implementation of general relativistic hydrodynamics in the moving-mesh \textsc{Arepo} code, including a dynamical spacetime evolution, provides a fundamentally new tool to simulate general relativistic problems in astrophysics.
\end{abstract}



\section{Introduction}\label{sec:Intro}

Numerical simulations are an important tool to study astrophysical systems involving compact objects such as core-collapse supernovae and binary mergers \citep{2012ARNPS..62..407J, 2012LRR....15....8F,2015IJMPD..2430012R,2017hsn..book.1053F,2017hsn..book.1671K,2017hsn..book.1605R,2017hsn..book.1095J,2017hsn..book.1805M,2017RPPh...80i6901B,2019JPhG...46k3002B,2019PrPNP.10903714B,2019RPPh...82a6902D,2019ARNPS..69...41S,2020GReGr..52..108B,2020ARNPS..70...95R,2021GReGr..53...27D,2022arXiv221207498J}.
The interpretation of the observations and the extraction of physics from such astronomical measurements to a large extent rely on numerical modelling of these events. For instance, observing binary neutron star (BNS) mergers provides the opportunity to study properties of high-density matter and the formation of heavy elements among several other fascinating aspects like short gamma-ray bursts. The recent simultaneous measurement of the inspiral stage of the BNS merger GW170817 and its electromagnetic counterparts \citep[][]{2017PhRvL.119p1101A,2017ApJ...848L..12A, 2017ApJ...851L..21V}, and especially the conclusions drawn from it, highlight the importance of numerical studies of these systems. Simulating compact objects can be challenging because many scenarios require the concurrent resolution of disparate length and time scales of a highly dynamical system in three dimensions and the inclusion of various physical effects. 

The bulk dynamics of such systems is governed by relativistic hydrodynamics in combination with a dynamical spacetime. There exist several approaches to numerically treat relativistic hydrodynamics: most prominent are Eulerian grid-based methods (including finite-difference, finite-volume or discontinuous Galerkin schemes) and Lagrangian smoothed particle hydrodynamics (SPH) \citep[for reviews see e.g.][and references therein]{wilson_mathews_2003, 2008LRR....11....7F, alcubierre2008introduction, baumgarte_shapiro_2010, rezzolla2013relativistic, 2015LRCA....1....1R, 2015LRCA....1....3M, shibata2015numerical}. \citet{2008LRR....11....7F}, \citet{2017RPPh...80i6901B} and \citet{2022arXiv220308139F} provide a survey of codes currently used to tackle general relativistic hydrodynamics (GRHD) problems mostly in the context of binary mergers; some recent relativistic SPH tools are presented in \citet{2019MNRAS.485..819L} (adopting a fixed metric) and in \citet{2021CQGra..38k5002R,2022EPJA...58...74D} (including a dynamical evolution of the spacetime and applications to neutron star mergers).

Both Eulerian grid-based methods and SPH have specific advantages and limitations (see the above references for more detailed discussions). Eulerian methods solve the GRHD equations on a static mesh, where many modern codes include (adaptive) mesh refinement techniques to resolve specific regions of interest. These methods accurately resolve shocks and fluid instabilities through the implementation of high-resolution shock-capturing methods. However, they may suffer from grid effects, require a special treatment of vacuum regions, and, in the case of compact object mergers, following small amounts of high velocity ejecta at large distances can be challenging.

SPH solves the Lagrangian hydrodynamics equations (comoving with the fluid) on particles representing a certain amount of rest mass. Advection and rest mass conservation are treated with high accuracy and the scheme offers an inherently adaptive resolution. Vacuum regions do not require a special treatment and tracer particles for nucleosynthesis calculations are trivially implemented. Traditionally, SPH is considered to resolve shocks and fluid instabilities poorer compared to high-resolution shock-capturing methods \citep[but see e.g.][and references therein for more modern techniques showing significant improvements]{2015LRCA....1....1R}. Notably, the Einstein equations cannot be solved in a particle-based discretization and thus require the inclusion of an additional computational grid and corresponding communication between both computational structures (similar problems may arise for treating other non-zero fields in the vacuum like magnetic fields).

\citet{2010MNRAS.401..791S} introduced the moving-mesh code \textsc{Arepo}, which combines some of the advantages of Lagrangian SPH and Eulerian mesh-based hydrodynamics. \textsc{Arepo} solves Newtonian hydrodynamics with a finite-volume approach on a moving unstructured mesh, which is constructed based on a set of mesh-generating points. The moving-mesh approach retains many of the advantages of mesh-based methods, while the mesh-generating points can move in an arbitrary way \citep[see][for more details]{2010MNRAS.401..791S}. Over the last years, \textsc{Arepo} has been employed for a wide range of astrophysical problems in cosmology, Type Ia supernovae, the common envelope phase in binary stars and various other systems \citep[see e.g.][]{2013ApJ...770L...8P, 2014Natur.509..177V, 2016ApJ...816L...9O, 2017MNRAS.470.4530W, 2019MNRAS.484.2047K, 2019Natur.574..211S, 2021A&A...649A.155G, 2022MNRAS.517.5260P}. A number of other moving-mesh codes have subsequently been developed and applied to various astrophysical problems \citep{2011ApJS..197...15D, 2012ApJ...755....7D, 2012ApJ...758..103G, 2013ApJ...775...87D, 2015ApJS..216...35Y, 2016A&C....16..109V, 2017MNRAS.471.3577C, 2022MNRAS.510.1315A} with several works investigating high-order schemes on moving meshes \citep[see e.g.][and references therein]{2013CCoPh..14..301D,Dumbser2017CentralWE,2020JCoPh.40709167G}. All these applications have generally shown the usefulness and benefits of the moving-mesh approach as compared to more traditional schemes.

Moving-mesh codes can follow the fluid motion and allow to flexibly place resolution in physically interesting regions. Hence, they offer adaptive resolution, which follows the matter motion including the possibility to split or merge cells and by this to adaptively increase or decrease the resolving power. The quasi-Lagrangian nature of the scheme reduces numerical advection errors. These elements make moving-mesh codes particularly interesting for simulating compact objects and in particular BNS systems. In recent years, some moving-mesh codes have been extended to include GRHD \citep{2017ApJ...835..199R, 2020MNRAS.496..206C}. However, all these implementations currently employ a fixed spacetime, and to date no moving-mesh code evolves the spacetime dynamically, as it would for instance be required to simulate neutron star mergers.

In this work we extend \textsc{Arepo} to simulate general relativistic systems (based on the upgraded implementation described in \citet{2016MNRAS.455.1134P}). We implement GRHD into the code employing the Valencia formulation \citep{1997ApJ...476..221B} and couple to it a solver for a dynamical spacetime. The Einstein equations are solved on an independent overlaid grid adopting the conformal flatness approximation \citep{1980grg1.conf...23I, 1996PhRvD..54.1317W}. We also include some additional modules to simulate neutron stars such as a high-density equation of state (EOS). We demonstrate the performance of the code in relativistic shock tube and blast wave problems. We further validate our implementation by computing equilibrium models of isolated neutron stars, which are benchmarked by comparing pulsation frequencies to perturbative results and other codes. Finally, we perform the first moving-mesh simulation of a BNS merger. We evolve the system for almost $40$~ms into the post-merger phase and discuss the dynamical properties of the remnant and the characteristics of the GW signal.

The paper is structured as follows. Sec.~\ref{sec:MainEqs} introduces the theoretical framework of our work. In Sec.~\ref{sec:Numerics} we provide details of our numerical implementation focusing on modifications with respect to the original code \citep{2010MNRAS.401..791S,2016MNRAS.455.1134P}. In Sec.~\ref{sec:TOVtests} we present simulations of isolated, static stars. In Sec.~\ref{sec:NSmergers} we describe the initial data for BNSs and present a BNS merger simulation. In the last section we provide a summary of our work and outline future plans. We also include appendices where we provide additional details on the theoretical formulation (Appendix~\ref{app:GWbackreaction}), present relativistic shock tube (Appendix~\ref{app:ShockTube}) and blast waves tests (Appendix~\ref{app:BlastWaves}), and investigate certain aspects of the numerical treatment with additional isolated neutron star and binary neutron star merger simulations (Appendix~\ref{app:NumSetup}). Throughout this work we set $c=G=1$, unless otherwise specified. Greek indices denote spacetime components, while Latin indices refer to spatial components.


\section{Theoretical formulation}\label{sec:MainEqs}
We briefly present the basic equations implemented in the relativistic version of \textsc{Arepo}. 

\subsection{Field equations}\label{subsec:GReqs}

We adopt the ADM formalism \citep{2008GReGr..40.1997A} to foliate the spacetime into a set of non-intersecting spacelike hypersurfaces with a constant coordinate time $t$. The general metric element then reads
\begin{equation}
 ds^2 = g_{\mu\nu} dx^\mu dx^\nu = \left( -\alpha^2 + \beta_i \beta^i \right) dt^2 + 2 \beta_i dx^i dt + \gamma_{ij} dx^i dx^j,
\end{equation}
where $g_{\mu\nu}$ is the spacetime $4-$metric, $\alpha$ denotes the lapse function, $\beta^i$ is the shift vector and $\gamma_{ij}$ the spatial $3-$metric.

In this work we impose the conformal flatness condition \citep{1980grg1.conf...23I, 1996PhRvD..54.1317W}, which approximates the spatial part of the metric as
\begin{equation}
 \gamma_{ij} = \psi^4 \hat{\gamma}_{ij},
\end{equation}
where $\psi$ is the conformal factor and $\hat{\gamma}_{ij}$ is the flat metric i.e.\ $\hat{\gamma}_{ij}=\delta_{ij}$ in Cartesian isotropic coordinates, which we use in our treatment.

Adopting the maximal slicing condition $\mathrm{Tr} (K_{ij})=0$, where $K_{ij}$ is the extrinsic curvature, the Einstein equations reduce to a set of five coupled nonlinear elliptic differential equations and read
\begin{align}
 \Delta\psi         =& -2\pi\psi^5 E - \frac{1}{8}\psi^5K_{ij}K^{ij}, \label{DEcfc:psi}\\
 \Delta(\alpha\psi) =& 2\pi\alpha\psi^5 (E+2S) + \frac{7}{8}\alpha\psi^5 K_{ij}K^{ij}, \label{DEcfc:alppsi}\\
 \Delta\beta^i      =& -\frac{1}{3}\partial^i\partial_j\beta^j 
                       + 2 \psi^{10} K^{ij} \partial_j \left(\frac{\alpha}{\psi^6}\right) 
                       +16\pi\alpha\psi^4 S^i \label{DEcfc:beta},
\end{align}
where $E$, $S$ and $S^i$ are matter sources terms. We adopt the energy-momentum tensor of a perfect fluid, namely
\begin{equation}
T^{\mu\nu} = \rho h u^\mu u^\nu + p g^{\mu\nu},
\end{equation}
where $\rho$ is the rest-mass density, $h=1+\epsilon+p/\rho$ the specific enthalpy, $\epsilon$ the specific internal energy, $p$ the pressure and $u^\mu$ the $4-$velocity of the fluid. Then, in the system of differential Eqs.~\eqref{DEcfc:psi}-\eqref{DEcfc:beta}, the various matter contributions in the source terms are given by
\begin{align}
E   &= \rho h\left(au^0\right)^2-p, \\
S   &= \rho h \left[\left(a u^0\right)^2-1\right] + 3p, \\
S^i &= \rho h \alpha u^0 u^\mu \gamma^i_\mu.
\end{align}

Within the conformal flatness approximation, the extrinsic curvature follows directly from the metric elements as
\begin{equation}
 K_{ij} = \frac{\psi^4}{2\alpha} \left( \delta_{ik}\partial_j\beta^k + \delta_{jk}\partial_i\beta^k - \frac{2}{3} \delta_{ij} \partial_k\beta^k \right).
\end{equation}

Following \citet{1998PhRvD..57.7299B} we introduce the definition $\beta^i=B^i-\frac{1}{4}\partial_i\chi$. Then Eq.~\eqref{DEcfc:beta} can be rewritten as two Poisson-like differential equations for the two auxiliary fields $B^i$ and $\chi$, which read
\begin{align}
 \Delta B^i   =& 2 \psi^{10} K^{ij} \partial_j \left(\frac{\alpha}{\psi^6}\right) 
                +16\pi\alpha\psi^4 S^i, \label{DEcfc:Beta} \\
 \Delta \chi  =& \partial_i B^i, \label{DEcfc:chi}
\end{align}
and can be solved iteratively.

For more details about the numerical implementation see Sec.~\ref{subsec:Metric} and \citet{2002PhRvD..65j3005O,2007A&A...467..395O}.

\subsection{General Relativistic Hydrodynamics}\label{subsec:GRHD}
The equations of GRHD result from the conservation laws for the energy-momentum tensor $T^{\mu\nu}$ and matter current density $J^\mu=\rho u^\mu$. By choosing a set of appropriate conserved variables, the conservation laws can be written in the form of a first-order flux-conservative hyperbolic system of equations which reads
\begin{equation}\label{GRHD:eqs}
 \partial_0 \Bigl( \sqrt{\gamma} \bmath{U}   \Bigr) + 
 \partial_i \Bigl( \sqrt{\gamma} \bmath{F}^i \Bigr) = \bmath{S},
\end{equation}
where $\bmath{U}$ is the state vector, $\bmath{F}^i$ the flux vector, $\bmath{S}$ is the source vector and $\gamma=\det(\gamma_{ij})$ the determinant of the 3-metric \citep{1997ApJ...476..221B,2008LRR....11....7F}.

The state, flux and source vectors are functions of the primitive variables $\bmath{W} = (\rho, \upsilon^i, \epsilon)$, where $\upsilon^i = (u^i/u^0+\beta^i)/\alpha$ is the fluid $3-$velocity. The state vector consists of the conserved variables and reads
\begin{equation} \label{stateU}
\bmath{U} = \left(
\begin{array}{c}
D\\
S_i\\
\tau
\end{array} 
\right) = 
 \left(
\begin{array}{c}
\rho W\\
\rho h W^2 \upsilon_i\\
\rho h W^2 - p - D
\end{array} 
\right),
\end{equation}
where $W=\alpha u^0=(1-\gamma_{ij}\upsilon^i\upsilon^j)^{-1/2}$ is the Lorentz factor. Furthermore, the flux and source vectors are given by
\begin{equation} 
\bmath{F}^i = \alpha \left(
\begin{array}{c}
D \left( \upsilon^i -\frac{\beta^i}{\alpha}\right) \\
S_j \left( \upsilon^i -\frac{\beta^i}{\alpha}\right) + p \delta^i_j \\
\tau \left( \upsilon^i -\frac{\beta^i}{\alpha}\right) + p \upsilon^i
\end{array} 
\right).
\end{equation} 
and 
\begin{equation} 
\bmath{S} = \alpha \sqrt{\gamma} \left(
\begin{array}{c}
0 \\
T^{\mu\nu} \left( \partial_\mu g_{\nu j} - \Gamma^\lambda_{\nu\mu} g_{\lambda j}\right) \\
\alpha \left( T^{\mu 0} \partial_\mu \ln{\alpha} - T^{\mu\nu} \Gamma^0_{\nu\mu} \right)
\end{array} 
\right),
\end{equation}
respectively. Here $\Gamma^\lambda_{\nu\mu}$ are the Christoffel symbols of the metric. In the following sections we also employ the definitions $\bmath{\mathcal{U}}=\sqrt{\gamma}\bmath{U}$ and $\bmath{\mathcal{F}}^i=\sqrt{\gamma}\bmath{F}^i$.

\subsection{Equation of state}\label{subsec:EoS}
In order to close the system of GRHD Eqs.~\eqref{GRHD:eqs} one needs to specify an EOS. We implement three different options for the EOS.

The first option is an (isentropic) polytropic EOS
\begin{align}
 p &= K\rho^\Gamma, \label{Eq:PresPoly} \\
 \epsilon &= \frac{K\rho^{\Gamma-1}}{(\Gamma-1)}, \label{Eq:EpsPoly}
\end{align}
where $K$ is the polytropic constant and $\Gamma$ is the polytropic index.

The polytropic EOS is suitable for an evolution of the system, where the equation for $\tau$ is not evolved. The value of the specific internal energy is instead analytically computed based on Eq.~\eqref{Eq:EpsPoly}.

An evolution with the polytropic EOS fails to capture a number of dynamical processes (e.g.\ shocks). Hence, we implement also an ideal gas EOS
\begin{equation}
 p=(\Gamma-1)\rho\epsilon,
\end{equation}
which we use for some of our tests (see Sec.~\ref{subsec:TOVCowling}).

Finally, we include a module for hybrid EOSs, which employs a zero-temperature tabulated microphysical EOS complemented by an ideal-gas component to capture thermal effects \citep{1993A&A...268..360J}. In this EOS, the pressure and specific internal energy read
\begin{align}
 p        &= p_{\mathrm{cold}}(\rho) + p_{\mathrm{th}}, \\
 \epsilon &= \epsilon_{\mathrm{cold}}(\rho) + \epsilon_{\mathrm{th}},
\end{align}
where $p_{\mathrm{cold}}$ and $\epsilon_{\mathrm{cold}}$ refer to the microphysical EOS and are functions of $\rho$. The thermal pressure is given by
\begin{equation}
 p_{\mathrm{th}} = (\Gamma_{\mathrm{th}}-1)\rho\epsilon_{\mathrm{th}},
\end{equation}
where $\epsilon_{\mathrm{th}}$ follows from $\epsilon_{\mathrm{th}}=\epsilon - \epsilon_{\mathrm{cold}}(\rho)$ and $\Gamma_{\mathrm{th}}$ is an appropriately chosen constant, typically in the range between $1.5$ and $2$ for neutron star applications \citep{2010PhRvD..82h4043B}. Within this framework, we estimate the temperature $T_{\mathrm{th}}$ based on the thermal energy through
\begin{equation}
 \epsilon_{\mathrm{th}}=\frac{kT_{\mathrm{th}}}{m_\mathrm{B}(\Gamma_\mathrm{th}-1)},
\end{equation}
where $k$ is the Boltzmann constant and $m_B$ the baryon mass.


\section{Numerical implementation}\label{sec:Numerics}
We describe the most important steps of our numerical implementation focusing on the modifications and additions to the original \textsc{Arepo} code \citep{2010MNRAS.401..791S,2016MNRAS.455.1134P}. This includes a number of standard methods, as well as additional techniques specific to the moving-mesh approach which we adopt for solving the GRHD equations. Based on the employed schemes, our implementation is formally second-order both in space and time.

\subsection{Time update}\label{subsec:TimeUpdate}
\textsc{Arepo} constructs an unstructured Voronoi mesh based on the positions of a set of mesh-generating points. The equations of hydrodynamics are discretized on this mesh in a finite-volume fashion \citep{2010MNRAS.401..791S}. Mesh-generating points can be moved simultaneously to the hydrodynamical evolution and this allows a dynamical reconfiguration of the computational grid. For each cell $i$ the volume-integrated conserved variables read
\begin{equation}
 \bmath{Q}_i = \int_{V_i} \bmath{\mathcal{U}} dV.
\end{equation}
The state $\bmath{Q}_i^n$ at time $t^n$ is evolved to the next timestep $t^{n+1}$ using Heun's method, which is a second-order Runge--Kutta scheme \citep{2016MNRAS.455.1134P}. The time-updated state $\bmath{Q}_i^{n+1}$ is given by
\begin{align}\label{Qupdate}
 \bmath{Q}_i^{n+1} =  \bmath{Q}_i^n
                    & - \frac{\Delta t}{2} \left( 
                      \sum_j A_{ij}^n     \hat{\bmath{F}}^n_{ij}(\bmath{W}_{ij}^n              ,\bmath{W}_{ji}^n) +
                      \sum_j A_{ij}^{\prime} \hat{\bmath{F}}^{\prime}_{ij}(\bmath{W}_{ij}^{\prime},\bmath{W}_{ji}^{\prime})
                      \right) \notag \\
                    & + \frac{\Delta t}{2} \left( 
                      \bmath{\mathcal{S}}_i^n + \bmath{\mathcal{S}}_i^{\prime}
                      \right),
\end{align}
where the index $j$ runs over all neighbouring cells of cell $i$ and $\Delta t$ is the timestep. $A_{ij}$ is the interface area between cells $i$ and $j$, while $\hat{\bmath{F}}_{ij}$ is an approximate Riemann solver estimate for the fluxes through $A_{ij}$ (see Sec.~\ref{subsec:RiemannSolver}). The fluxes depend on $\bmath{W}_{ij}$ and $\bmath{W}_{ji}$, which are the reconstructed primitive variables from the center of cell $i$ (or $j$ respectively) to the cell interfaces (see Sec.~\ref{subsec:Recon}). $\bmath{\mathcal{S}}_i = \int_{V_i} \bmath{S} dV$ are the volume-integrated source terms computed for cell $i$.

Within the Heun method a forward Euler integration has to be performed, which estimates the states at the end of the timestep as
\begin{equation}
 \bmath{Q}_i^{\prime} =  \bmath{Q}_i^n
                      - \Delta t \sum_j A_{ij}^n 
                      \hat{\bmath{F}}^n_{ij}(\bmath{W}_{ij}^n, \bmath{W}_{ji}^n)
                      + \Delta t \bmath{\mathcal{S}}_i^n.
\end{equation}
These estimates are used to compute the fluxes $\hat{\bmath{F}}^{\prime}_{ij}$ and source terms $\bmath{\mathcal{S}}_i^{\prime}$ at the end of the timestep, where the primitives are recovered and reconstructed from $ \bmath{Q}_i^{\prime}$.

Within the timestep we also move the mesh-generating points and construct a new mesh. As a result, the mesh geometry is different at the beginning and the end of the timestep. This is already apparent in Eq.~\eqref{Qupdate}, where we employ different terms for the face areas at the beginning and end of the timestep i.e. $A_{ij}^n$ and $A_{ij}^{\prime}$ respectively. We update the positions of the mesh-generating points as
\begin{equation}
 \bmath{r}_i^{n+1} = \bmath{r}_i^{n} + \frac{\Delta t}{2} \left( \bmath{w}_i^n+\bmath{w}_i^\prime \right)
 = \bmath{r}_i^{n} + \Delta t \bmath{w}_i^n,
\end{equation}
where $\bmath{r}_i$ denotes the coordinates of the mesh-generating point and $\bmath{w}_i$ is the point's velocity. As described in \citet{2016MNRAS.455.1134P}, we keep the velocity of each mesh-generating point constant throughout the whole timestep ( i.e.\ $\bmath{w}^{\prime}_i=\bmath{w}^n_i$). By doing so, the mesh which is constructed at the end of the current timestep matches the mesh at the beginning of the next timestep (i.e.\ $A_{ij}^{\prime}\equiv A_{ij}^{n+1}$). This highlights the benefit of using Heun's method because it requires practically only one mesh construction per timestep.

In the applications discussed in this paper the metric fields do not change too rapidly, in the sense that their variations over a timestep are small. In this situation one may avoid solving the metric field equations in every timestep to save computing time. We can explicitly solve the field equations every few timesteps and use this information to extrapolate the metric in the intermediate timesteps. We note that similar approaches are used also in other codes which employ the conformal flatness condition \citep[see e.g.][]{2002A&A...388..917D,2011A&A...528A.101B,2020CQGra..37n5015C,2021MNRAS.508.2279C}. In the simulations performed in this work, we update the metric at the beginning of each of the two substeps of Heun's method. We explicitly solve the metric field equations in each Heun substep for the first nine timesteps. Subsequently, we call the metric solver in the first Heun substep of every fifth timestep. For the remaining substeps we estimate the metric using a parabolic extrapolation based on the last three metric solutions. The extrapolation is performed on the metric grid. Explicitly solving the metric fields equations in the first timesteps ensures the stability of the scheme after importing initial data and provides the necessary number of collocation points for the extrapolation. We test this procedure and evaluate the agreement with simulations where we solve the metric equations in every substep in Appendix \ref{subapp:NumSetupFreq}. The extrapolation significantly reduces the computational effort.

\textsc{Arepo} can update cell states based on an individual time step for each cell. It employs a power-of-two hierarchy to account for different cell sizes and achieve synchronization \citep[see][for more details]{2010MNRAS.401..791S}. We have not yet experimented with this feature, which we plan to test in future work. Hence, at the moment, we do not employ this functionality and use a single global time step instead. In all the simulations presented in this work we apply the Courant–Friedrichs–Lewy (CFL) condition with a CFL factor $C_{\mathrm{CFL}}=0.3$ to compute the maximum allowed time step $\Delta t_i$ for each cell. For a cell with volume $V_i$ we employ $[3V_i/(4\pi)]^{1/3}$ as an effective radius for the cell in order to compute $\Delta t_i$. Then the global time step is given by 
\begin{equation}
\Delta t = \frac{T_\mathrm{tot}}{2^N},
\end{equation}
where $T_\mathrm{tot}$ is the total simulation time, which is a free parameter, and $N$ is the smallest integer value for which $\Delta t < \min\limits_i \Delta t_i$ holds.

\subsection{Reconstruction of primitive variables}\label{subsec:Recon}
To compute the flux terms in Eq.~\eqref{Qupdate}, we need to reconstruct the primitive variables from the cell center to the mid-points of the faces. \textsc{Arepo} linearly approximates any quantity $\phi$ from the center of mass of the cell $\bmath{s_i}$ to any other point within the cell $\bmath{r}$ as
\begin{equation}
 \phi(\bmath{r}) = \phi(\bmath{s}_i) + \langle \nabla \phi \rangle_i \cdot (\bmath{r} - \bmath{s}_i),
\end{equation}
where $\langle \nabla \phi \rangle_i$ is an estimate of the gradient of $\phi$ within the cell \citep[see][for more details on computing the gradient estimate]{2016MNRAS.455.1134P}.

In addition, the gradients are slope-limited. The original implementation of \textsc{Arepo} replaces the gradient estimate $\langle \nabla \phi \rangle_i$ by $\alpha_i \langle \nabla \phi \rangle_i$, where $\alpha_i = \min (1,\psi_{ij})$ \citep{2010MNRAS.401..791S}. The index $j$ refers to neighbouring cells of cell $i$ and the quantity $\psi_{ij}$ is defined as
\begin{equation}\label{ReconOrig}
\psi_{ij} = \left\{
\begin{array}{cl}
(\phi_i^\mathrm{max}-\phi_i)/\Delta\phi_{ij} &, \Delta\phi_{ij}>0 \\
(\phi_i^\mathrm{min}-\phi_i)/\Delta\phi_{ij} &, \Delta\phi_{ij}<0 \\
1 &, \Delta\phi_{ij}=0
\end{array}
\right.,
\end{equation}
where $\Delta\phi_{ij} = \langle \nabla \phi \rangle_i \cdot (\bmath{f}_{ij} - \bmath{s}_i)$ is the estimate of the change in $\phi$ between the center of cell $i$ and the centroid of the interface between cells $i$ and $j$ (denoted by $\bmath{f}_{ij}$) \citep{BARTH1989}. Furthermore, $\phi_i^\mathrm{max}$ and $\phi_i^\mathrm{min}$ are the maximum and minimum values among all the neighbours, including cell $i$.

Alternatively, we apply the monotonized-central difference (MC) slope limiter \citep{VANLEER1977276} to the gradient estimate in order to ensure that the scheme is total variation diminishing \citep{1983JCoPh..49..357H,1984SJNA...21....1H}. We follow the approach outlined in \citet{DARWISH2003599} for the extension of slope limiters to unstructured grids. Other choices for the slope limiter are also possible. However, the MC slope limiter was shown to perform better in simulations of single relativistic stars \citep{2002PhRvD..65h4024F}.

For the simulations that we present in this work, we employ slope-limited reconstruction with the MC slope limiter, unless otherwise stated. We also highlight that there is ongoing work on higher-order reconstruction schemes on moving meshes \citep[see e.g.][and references therein]{Dumbser2017CentralWE,2020JCoPh.40709167G}.

\subsection{Riemann problem}\label{subsec:RiemannSolver}
The original (Newtonian) implementation of \textsc{Arepo} solves the Riemann problem at each face in the rest-frame of the face. This involves estimating the velocity $\tilde{\bmath{w}}_{ij}$ of the common face between each pair of neighbouring cells $i$ and $j$ \citep[see Sec.~3.3 in][for how to estimate $\tilde{\bmath{w}}_{ij}$]{2010MNRAS.401..791S} and boosting the corresponding cell states by $\tilde{\bmath{w}}_{ij}$. The states on both sides of the mid-point of the face (denoted as left/right) follow from reconstructing the primitive variables. In order to apply an approximate $1\mathrm{D}$ Riemann solver, the left/right states need to be rotated such that the $x-$axis aligns with the normal vector of the face. The solution of the Riemann problem follows from sampling the self-similar solution along $x/t=0$. The solution is then rotated and boosted back to the initial ``lab'' frame and used to compute the flux terms in Eq.~\eqref{Qupdate}.

For simplicity, in our general relativistic treatment we do not perform the exact same steps. Instead, we follow a different methodology introduced in \citet{2011ApJS..197...15D} and solve the Riemann problem in the ``lab'' frame. In particular, we employ the HLLE solver \citep{10.2307/2030019,1988SJNA...25..294E} and sample the solution along $x/t=\tilde{\bmath{w}} \cdot \hat{\eta}$ to capture the correct HLLE state, where $\hat{\eta}$ is the (outward) normal vector to the face. The numerical fluxes, which enter Eq.~\eqref{Qupdate}, then read
\begin{equation}\label{Eq:Fluxlab}
\hat{\bmath{F}}_{ij} = \bmath{\mathcal{F}}_{ij}^{1\mathrm{D}} - \tilde{\bmath{w}}_{ij} \cdot \hat{\eta} \bmath{\mathcal{U}}_{ij}^{1\mathrm{D}},
\end{equation}
where $\bmath{\mathcal{F}}_{ij}^{1\mathrm{D}}$ and $\bmath{\mathcal{U}}_{ij}^{1\mathrm{D}}$ are computed by the HLLE solver in the ``lab'' frame. The second term in Eq.~\eqref{Eq:Fluxlab} accounts for advection by the moving face.

\subsection{Conversion from conserved to primitive variables}\label{subsec:ConsToPrim}
It is evident from Eq.~\eqref{Qupdate} that at the end of each timestep we know the volume-integrated conserved variables $\bmath{Q}$, or in turn the conserved variables $\bmath{U}$. To solve the GRHD equations one needs to compute quantities (e.g.\ the fluxes $\bmath{F}^i$ and sources $\bmath{S}$) which require the primitive variables. While $\bmath{U}$ analytically follows from the primitive variables, obtaining the primitive from the conserved variables requires a numerical solution. The recovery of primitive variables is a common intricate task of GRHD schemes.

We employ a widely used and tested method \citep[see e.g.][]{rezzolla2013relativistic}, which is based on a Newton--Raphson scheme. As a first step, we express the density and specific internal energy as
\begin{align}
 \rho     &= \frac{D\sqrt{Q^2-S^2}}{Q}, \\
 \epsilon &= \left(\sqrt{Q^2-S^2} - \frac{pQ}{\sqrt{Q^2-S^2}} - D\right) \Bigg/ D,
\end{align}
based on Eqs.~\eqref{stateU} and the definitions $S^2=\gamma^{ij} S_iS_j$ and $Q=\tau+p+D$. Then, for a generic EOS $p=p(\rho,\epsilon)$, we employ a Newton--Raphson method to solve the equation
\begin{equation}
 p - \hat{p}\left[\rho\left(\bmath{U},p\right),\epsilon\left(\bmath{U},p\right)\right] = 0,
\end{equation}
starting from an initial guess for the pressure $p$ (e.g. the pressure at the cell center in the previous timestep for accelerated root-finding). We compute the necessary derivatives $\partial \hat{p}/\partial\rho$ and $\partial \hat{p}/\partial\epsilon$ numerically. As an additional measure we reset the primitive variables to atmosphere values if for a cell $p<0$ or the density $\rho$ is below a threshold value $\rho_\mathrm{thr}$ (see Sec.~\ref{subsec:AddDetails} for details). The conserved variables are then recomputed for the new primitives.

\subsection{Mesh geometry}\label{subsec:MeshGeometry}
Arguably two of the most important aspects in a moving-mesh simulation are the initial positions of the mesh-generating points and how the points move during the simulation. The initial distribution of the points determines the initial geometry of the mesh, while point motion determines how the mesh geometry evolves. A mesh which is well-adapted to the geometrical and physical aspects of the problem at hand captures the physics more accurately even with fewer cells, since resolution is distributed more appropriately in the simulation domain.

In the various tests that we present in the following sections, we use different initial mesh-generating point distributions for different tests. Furthermore, we perform both moving-mesh, as well as static-mesh simulations, i.e.\ calculations where the cells move or remain fixed at their initial positions, respectively. In our moving-mesh simulations each point moves with the local fluid coordinate velocity with possibly a small correction to this velocity to ensure that the mesh does not become too irregular \citep[see Sec.~4 in][for more details]{2010MNRAS.401..791S}. For the mesh regularity we adopt a more recent criterion proposed in \citet{2012MNRAS.425.3024V} \citep[see][for a summary]{2020ApJS..248...32W}. For each cell $i$, we define
\begin{equation}
\alpha_\mathrm{max}=\max\limits_j\left({\sqrt{A_j/\pi}/h_j}\right)
\end{equation}
to estimate how round the cell is based on the area of each face $A_j$ and its distance from the mesh-generating point $h_j$. We identify cells which satisfy $\alpha_\mathrm{max}>0.75\beta$ as irregular, where $\beta$ is a free parameter typically set to $2.25$. For irregular cells we include a corrective velocity component to the motion of the mesh-generating point, which drifts the point closer to the center of mass of the respective cell. The corrective velocity reads
\begin{align}\label{Eq:RegVel}
 \bmath{v}_\mathrm{corr} = \left\{
\begin{array}{ll}
      0                                                                                             & ,\alpha_\mathrm{max} \leq 0.75\beta \\
      f_\mathrm{shaping} \frac{\bmath{s}_i-\bmath{r}_i}{d_i} \frac{\alpha_\mathrm{max}-0.75\beta}{0.25\beta} v_\mathrm{char} & ,0.75\beta<\alpha_\mathrm{max} \leq \beta\\
      f_\mathrm{shaping} \frac{\bmath{s}_i-\bmath{r}_i}{d_i} v_\mathrm{char}                                                  & ,\alpha_\mathrm{max} > \beta \\
\end{array} 
\right. ,
\end{align}
where $d_i$ is the distance between the center of mass ($\bmath{s}_i$) and the mesh-generating point ($\bmath{r}_i$). We consider a fraction $f_\mathrm{shaping}$ (typically $0.5$) of the characteristic speed $v_\mathrm{char}$, while we set $v_\mathrm{char}$ to the sound speed in the cell.

Distributing the mesh-generating points carefully and allowing them to follow the fluid motion ensures that resolution is focused on the physically interesting regions and minimizes advection across cells. However, some problems might additionally benefit from increasing or decreasing the resolution locally in a flexible way. \textsc{Arepo} allows for cell refinement or derefinement based on nearly arbitrary criteria \citep[see Sec.~6 in][]{2010MNRAS.401..791S}. Different criteria can be employed to dynamically change the local geometry of the mesh, effectively adding resolution where it is needed and reducing the resolution where it is deemed redundant.

We provide details regarding the initial mesh geometry and whether we enable cell refinement/derefinement in the discussion of each test, since the choices strongly depend on the concrete application. We emphasize that a direct comparison between moving-mesh and static-mesh simulations is not necessarily straightforward, even in cases where the initial meshes are identical in moving-mesh and static-mesh simulations. The reason is that in moving-mesh calculations the cells rearrange over time. Hence the mesh can in principle evolve to a setup which differs significantly from the initial geometry.

\subsection{Additional code details}\label{subsec:AddDetails}
Grid-based hydrodynamics approaches require a special treatment for vacuum regions. We employ an artificial atmosphere, i.e. we place cells with a very low density $\rho_\mathrm{atm}$ in vacuum regions. During the evolution the numerical treatment resets the density to this value if the density of a cell falls below a threshold value $\rho_\mathrm{thr}$. In the tests which we present in the following sections we set $\rho_\mathrm{thr}=10\times\rho_\mathrm{atm}$ and $\rho_\mathrm{atm} = (10^{-7}-10^{-8})\times \rho_{\mathrm{max}}(t)$, where $\rho_{\mathrm{max}}(t)$ is the maximum density throughout the whole domain at any given time $t$. Hence, the atmosphere density changes in time if the maximum density oscillates. This criterion captures cells outside the neutron stars, where one should formally have vacuum. In these cells we also set the velocities to zero, while the pressure and specific internal energy follow from a polytropic EOS with $K=100$ and $\Gamma=2$. Subsequently, we update the conserved variables in these atmospheric cells based on the new set of primitives. We note that the values of $\rho_\mathrm{atm}$ and $\rho_\mathrm{thr}$ can be adjusted based on the aspects of the problem at hand. For instance, to follow BNS merger ejecta, lower atmosphere values are desirable, which we will explore in future work.

Formally, we adopt periodic boundary conditions for the hydro mesh in our calculations. We emphasize that this choice does not affect the evolution of the system because the outer boundaries are placed far away from the regions of physical interest, where all the cells have densities below the atmosphere threshold throughout the whole simulation (i.e. no matter reaches the boundaries). The size of the numerical domain varies in different tests to cover the whole physical system. As said, the exact size does not play any role since we ensure that in all simulations the physical domain of interest is surrounded by atmosphere cells. Hence, the size of the numerical domain can in principle be chosen arbitrarily large.

If the numerical domain is chosen to be very large compared to the physical system under consideration, we would typically fill the outer parts of the numerical domain with increasingly larger cells, i.e. by placing the mesh generating points more sparsely, to minimize the computational overhead. The chosen configuration should not lead to an abortion of the mesh construction algorithm, but we did not face any issues in this regard.

\subsection{Solution of field equations}\label{subsec:Metric}
We solve the metric field Eqs.~\eqref{DEcfc:psi}, \eqref{DEcfc:alppsi}, \eqref{DEcfc:Beta} and \eqref{DEcfc:chi} on an independent uniform Cartesian grid. We employ a multigrid algorithm \citep[see e.g.][]{10.5555/357695} and solve the differential equations iteratively until they converge. Boundary conditions for the solution of Eqs.~\eqref{DEcfc:psi} and \eqref{DEcfc:alppsi} are computed based on a multipole expansion of the fields. Formally, we can write the solution to Eq.~\eqref{DEcfc:psi} at a point with coordinates $\bmath{r}$ as
\begin{equation}\label{Eq:MultipoleExp}
\Delta\psi = S_\psi \Rightarrow \psi(\bmath{r}) = -\frac{1}{4\pi} \int \frac{S_\psi(\bmath{r}^\prime)}{|\bmath{r}-\bmath{r}^\prime|} d^3\bmath{r}^\prime,
\end{equation}
where $S_\psi$ collectively refers to the terms on the right-hand side of Eq.~\eqref{DEcfc:psi} and we integrate over the metric grid ($\bmath{r}^\prime$ is a coordinate vector) \citep[e.g.][]{2007A&A...467..395O}. Then, boundary conditions for Eq.~\eqref{DEcfc:psi} follow from expanding Eq.~\eqref{Eq:MultipoleExp} up to quadrupole order. We note that we consider only the monopole contribution from the non-compact term in $S_\psi$ (i.e. the term involving $K_{ij}K^{ij}$). We follow a similar procedure to compute boundary conditions for Eq.~\eqref{DEcfc:alppsi}.

We impose fall-off boundary conditions in order to solve Eqs.~\eqref{DEcfc:Beta} and \eqref{DEcfc:chi}, namely we approximate the fields $B^i$, $\chi$ at a point $\bmath{r}$ (e.g. at the boundaries) as
\begin{align}
B^i(\bmath{r})   &= \frac{b^i(\bmath{r})}{b^i(\bmath{r}_p)} B^i(\bmath{r}_p), \\
\chi(\bmath{r})  &= \frac{c(\bmath{r})}{c(\bmath{r}_p)} \chi^i(\bmath{r}_p).
\end{align}
where $\bmath{r}_p$ is the projection of $\bmath{r}$ on the grid boundary. Here, $b^i$ and $c$ capture the lowest-order fall-off behavior of the respective fields and read $b^x = y/r^3$, $b^y = x/r^3$, $b^z = xyz/r^7$ and $c = xy/r^5$ in the employed coordinates \citep[see][but note the different coordinate system]{1998PhRvD..57.7299B}.

The metric solver implementation originates from \citet{2002PhRvD..65j3005O}, where they provide more details. The grid size is chosen such that it covers the physical domain of interest well. The metric grid can be smaller than the hydro grid. For the BNS simulations discussed below, the metric grid will well cover the orbit of the binary but matter can extend beyond the metric grid (e.g.~ejecta) and the boundaries of the hydro grid are much farther out. Beyond the metric grid we employ the same treatment of the metric fields as on the boundaries of the metric grid using the aforementioned expansion and fall-off conditions. With the treatment beyond the metric grid being consistent with the boundary conditions of the metric grid, we do not notice any spurious effects when matter leaves the domain of the metric grid.
 
In our implementation hydrodynamic quantities and metric potentials are solved on different grids. To solve the GRHD equations we need to interpolate the metric fields to various positions e.g.\ the mesh-generating point positions, the center of mass of Voronoi cells or the mid-point of the interfaces between neighbouring cells. Vice versa, solving the metric field equations requires knowledge of the hydrodynamic variables at the positions of the metric grid points. We perform the mappings in the following way:
\begin{enumerate}
  \item \textit{Metric grid to hydro mesh}: Knowing the metric fields on a uniform Cartesian grid, we interpolate to any point if it lies within the metric grid. We interpolate the metric fields with a $3^{\mathrm{rd}}$ order Lagrange polynomial. We cannot apply the same approach to compute the metric fields at points outside the metric grid domain. Instead, we employ the multipole expansions and fall-off conditions that we used to compute the metric boundary conditions for these distant cells.
  
  \item \textit{Hydro mesh to metric grid}: The inverse mapping is significantly more complicated because the hydrodynamics mesh is unstructured. The main component is a tree walk \citep{2010MNRAS.401..791S} to locate the mesh-generating point which lies closest to each metric grid point. At the end of the tree search, each metric grid point is placed in a hydrodynamics cell. We then directly adopt all necessary hydrodynamical variables from the moving-mesh cell for the respective metric grid point.
  
  We have found this simple approach to be rather robust. As expected, its accuracy improves as the number of hydrodynamic cells increases. Furthermore, it benefits from the fact that physically important regions are more densely populated with mesh points. As a result, the distance between the centers of the moving-mesh cells and the metric grid points is typically quite small. As a way to monitor the accuracy of the approach, we compare the baryonic mass on the different grids for the simulations discussed in this work and find that they agree to at least $0.1\%$. The mass on the metric grid oscillates by this magnitude around the value of the unstructured hydro mesh without any systematic trend. The impact of this mismatch on for instance the phase evolution is certainly surpassed by the use of the conformal flatness approximation in the first place. We plan to further examine the mismatch in the future. Simple extensions (e.g.\ via the use of gradients $\langle \nabla \phi \rangle$ discussed in Section \ref{subsec:Recon}) can potentially prove more accurate without significantly adding to the computational cost.
\end{enumerate}

Finally, we remark that the treatment of a black hole requires certain modifications of the gravity solver within the conformal flatness approximation as in~\citet{2014ApJ...795L...9B,2015MNRAS.448..541J}, which we leave for future work.
 
\subsection{Gravitational wave backreaction and extraction}\label{subsec:Backreaction}
The conformal flatness approximation does not include gravitational radiation by construction which, however, can be important in some applications like for instance BNSs. Therefore, we complement our approach by adding a backreaction scheme to emulate gravitational wave energy and angular momentum losses.

We closely follow the implementation of \citet{2007A&A...467..395O}, which consists of adding a small, non-conformally flat correction to the metric based on a post-Newtonian analysis presented in \citet{2003PhRvD..68h4001F} \citep[see also][]{1990MNRAS.242..289B}. We outline the formalism in Appendix \ref{app:GWbackreaction}. For neutron star merger simulations this approach shows generally a good agreement with fully general relativistic simulations comparing for instance the post-merger gravitational wave emission, black-hole formation or ejecta and torus masses \citep[see e.g.][]{2012PhRvD..86f3001B,2013ApJ...773...78B,2021PhRvD.103l3004B,2022PhRvD.106d4026K}.


\section{Tolman-Oppenheimer-Volkoff star}\label{sec:TOVtests}
We present shock tube tests and relativistic blast waves in Appendices~\ref{app:ShockTube} and~\ref{app:BlastWaves}, and focus in this section on the evolution of isolated neutron stars. We evolve a static equilibrium neutron star and extract its fundamental radial mode frequency to verify our implementation. For a first test, we evolve a neutron star described by a polytrope while keeping the metric fixed (Cowling approximation). This analysis targets our general relativistic hydrodynamics implementation alone. As a second test, we evolve a neutron star described by a microphysical EOS, while dynamically evolving the spacetime as well. This setup tests our GRHD implementation, our metric solver and their coupling, as well as our microphysics modules.

Table~\ref{tab:example} summarizes the features and parameters of all simulations discussed in this paper including the BNS merger runs described in Sec.~\ref{sec:NSmergers} and Appendix~\ref{app:NumSetup}. 

\begin{table*}
\caption{Summary of the simulations presented in this study. The first column specifies the type of system simulated. Second column denotes if the spacetime was fixed or evolved dynamically. Third column indicates if the mesh was moving during the simulation. The fourth column contains information about the symmetry of the initial grids, which we employ in the different simulations. Fifth column indicates whether we employ cell refinement and derefinement in the respective simulation. In the sixth column we provide an estimate for the resolution. In moving-mesh simulations the resolution changes dynamically (see main text for more details on each simulation). In the case of the BNS systems the resolution refers to the high-density regime in the post-merger phase. Columns seven and eight list the EOS and masses of the systems. Finally, the last column reports the characteristic frequency extracted in every simulation. In the case of TOV stars this refers to the frequency of the radial mode ($f_\mathrm{0}$), while for the BNS systems it is the dominant frequency in the post-merger phase. The seventeenth row ``pert. TOV star'' provides the perturbative result of the isolated TOV star for comparison. The TOV simulations marked with ${}^\ast$ and ${}^\dag$ refer to the simulation with \textsc{Arepo}'s standard slope-limited gradient and the run with the regularization parameters set to $(f_\mathrm{shaping},\beta)=(0.5,3)$, respectively. The systems marked with ${}^\ddag$ refer to the simulations discussed in Appendix~\ref{subapp:NumSetupFreq}. For the BNS simulation labelled with ${}^\ddag$ we do not provide a characteristic frequency, because we evolve the system for $<5~\mathrm{ms}$ in the post-merger phase. For simulations labeled ``case (i)'', ``case (ii)'' and ``case (iii)'' see Appendix~\ref{subapp:NumSetupRes}.}
\label{tab:example}
\resizebox{\textwidth}{!}{%
\begin{tabular}{ccccccccc}
\hline
System & Spacetime & Mesh & Hydro grid setup & Cell refinement/ & Resolution & EOS & Gravitational  & Characteristic \\
& & motion & (region around stars) & derefinement & [m] & & Mass [$M_{\odot}$] & Frequency \\
\hline
TOV star & Fixed & Moving & Uniform Cartesian & No & $\approx221.4$   & Ideal gas & 1.4  & $f_\mathrm{0}=2.664$~kHz \\
TOV star & Fixed & Moving & Uniform Cartesian & No & $\approx184.5$   & Ideal gas & 1.4  & $f_\mathrm{0}=2.668$~kHz \\
TOV star   & Fixed & Moving & Uniform Cartesian & No & $\approx147.6$   & Ideal gas & 1.4  & $f_\mathrm{0}=2.672$~kHz \\
TOV star & Fixed & Moving & Uniform Cartesian & No & $\approx110.7$   & Ideal gas & 1.4  & $f_\mathrm{0}=2.674$~kHz \\
TOV star${}^\ast$ & Fixed & Moving & Uniform Cartesian & No & $\approx147.6$   & Ideal gas & 1.4  & $f_\mathrm{0}=2.677$~kHz \\
TOV star${}^\dag$ & Fixed & Moving & Uniform Cartesian & No & $\approx147.6$   & Ideal gas & 1.4  & $f_\mathrm{0}=2.672$~kHz \\
TOV star & Fixed & Moving & Spherical & No & $\approx191$   & Ideal gas & 1.4  & $f_\mathrm{0}=2.661$~kHz \\
TOV star   & Fixed & Static & Uniform Cartesian & No & $\approx147.6$   & Ideal gas & 1.4  & $f_\mathrm{0}=2.682$~kHz \\
TOV star   & Fixed & Static & Cartesian (lower crust resolution) & No & $\approx147.6$ \& $295.3$   & Ideal gas & 1.4  & $f_\mathrm{0}=2.674$~kHz \\
TOV star   & Dynamical & Moving & Spherical & No & $\approx253.7$ & H4 + $\Gamma_\mathrm{th}=1.75$         & 1.41 & $f_\mathrm{0}=2.318$~kHz \\
TOV star${}^\ddag$   & Dynamical & Moving & Spherical & No & $\approx253.7$ & H4 + $\Gamma_\mathrm{th}=1.75$         & 1.41 & $f_\mathrm{0}=2.316$~kHz \\
TOV star   & Dynamical & Moving & Spherical & No & $\approx191$ & H4 + $\Gamma_\mathrm{th}=1.75$         & 1.41 & $f_\mathrm{0}=2.343$~kHz \\
TOV star${}^\ddag$   & Dynamical & Moving & Spherical & No & $\approx191$ & H4 + $\Gamma_\mathrm{th}=1.75$         & 1.41 & $f_\mathrm{0}=2.344$~kHz \\
TOV star   & Dynamical & Moving & Spherical + Random & No & $\approx191$ & H4 + $\Gamma_\mathrm{th}=1.75$         & 1.41 & $f_\mathrm{0}=2.349$~kHz \\
TOV star   & Dynamical & Moving & Spherical & No & $\approx162.4$ & H4 + $\Gamma_\mathrm{th}=1.75$         & 1.41 & $f_\mathrm{0}=2.352$~kHz \\
TOV star   & Dynamical & Static & Spherical & No & $\approx191$ & H4 + $\Gamma_\mathrm{th}=1.75$         & 1.41 & $f_\mathrm{0}=2.358$~kHz \\
pert. TOV star   & Dynamical & -- & -- & -- & -- & H4          & 1.41 & $f_\mathrm{0}=2.385$~kHz \\
BNS merger & Dynamical & Moving & \begin{tabular}{@{}c@{}} Spherical (based \\on mass distribution) \end{tabular}  & Yes & $\approx162$     & DD2 + $\Gamma_\mathrm{th}=1.75$        & 1.35+1.35  & $f_\mathrm{peak}=2.56$~kHz \\
\begin{tabular}{@{}c@{}} BNS merger \\(case (i)) \end{tabular} & Dynamical & Moving & \begin{tabular}{@{}c@{}} Spherical (based \\on mass distribution) \end{tabular}  & Yes & $\approx182$     & DD2 + $\Gamma_\mathrm{th}=1.75$        & 1.35+1.35  & $f_\mathrm{peak}=2.55$~kHz \\
\begin{tabular}{@{}c@{}} BNS merger \\(case (ii)) \end{tabular} & Dynamical & Moving & \begin{tabular}{@{}c@{}} Spherical (based \\on mass distribution) \end{tabular}  & Yes & $\approx162$     & DD2 + $\Gamma_\mathrm{th}=1.75$        & 1.35+1.35  & $f_\mathrm{peak}=2.56$~kHz \\
\begin{tabular}{@{}c@{}} BNS merger \\(case (iii)) \end{tabular} & Dynamical & Moving & \begin{tabular}{@{}c@{}} Spherical (based \\on mass distribution) \end{tabular}  & Yes & $\approx155$     & DD2 + $\Gamma_\mathrm{th}=1.75$        & 1.35+1.35  & $f_\mathrm{peak}=2.57$~kHz \\
BNS merger${}^\ddag$ & Dynamical & Moving & \begin{tabular}{@{}c@{}} Spherical (based \\on mass distribution) \end{tabular}  & Yes & $\approx170$     & DD2 + $\Gamma_\mathrm{th}=1.75$        & 1.35+1.35  & -- \\
\hline
\end{tabular}
}
\end{table*}

\subsection{Cowling approximation}\label{subsec:TOVCowling}

\subsubsection{Initial data}\label{subsec:TOVid}
We solve the Tolman-Oppenheimer-Volkoff (TOV) equations and compute a $1.4~M_\odot$ polytropic neutron star with $K=100$ and $\Gamma=2$ (central density $\rho_\mathrm{c}=1.28\times10^{-3}$ in $c=G=1$ units). This stellar model is a common choice and allows us to compare our evolutions within the Cowling approximation with results from previous works \citep{2002PhRvD..65h4024F}.

We map the primitive quantities from the 1D TOV solution to a mesh-generating point distribution which is used to construct a Voronoi mesh. In this simulation the hydrodynamical simulation domain is a cube with side length $58~M_\odot\approx85.6$~km, hence significantly larger than the stellar radius ($R\approx12$km in isotropic coordinates). We set up the mesh-generating points to obtain a uniform Cartesian grid with high resolution in the center of the computational domain to cover the star. This particular mesh setup allows us to compare directly to \citet{2002PhRvD..65h4024F}, where also a Cartesian grid is employed. The central, high-resolution mesh is a cube with side length $24~M_\odot\approx35.4$~km and a cell size $h=0.1~M_\odot\approx147.6$m. We cover the rest of the simulation domain with points which lead to a low resolution mesh. Throughout the whole evolution these outer parts of the computational domain are atmosphere cells and thus the exact point distribution is irrelevant. Even a very low number of mesh-generating points is already sufficient (in the case of this particular setup $0.2\%$ of all cells), provided the mesh-construction algorithm can create a mesh.

We excite the radial mode by adding a radial $3$-velocity perturbation of the form
\begin{equation}\label{eq:RadPert}
 \delta \upsilon_r = A \sin{\left( \frac{\pi r}{R} \right)},
\end{equation}
where $A=-0.005$, $r$ is the radial distance from the stellar center and $R$ is the stellar radius \citep[see e.g.][]{2006MNRAS.368.1609D}.

In our Cowling tests we compute the metric fields at any point in our simulation domain (e.g.\ mesh-generating point positions, centers of mass of the hydrodynamic cells) by interpolating the high-resolution metric function profiles which we obtain from our TOV solution. We set the atmosphere density to $\rho_\mathrm{atm}=10^{-8}\times\rho_\mathrm{max}$ and consider any cell with $\rho<10\times\rho_\mathrm{atm}$ to be part of the atmosphere. We evolve the polytropic initial data with an ideal gas EOS and thus also evolve the energy equation.

\subsubsection{Simulations}\label{subsec:TOVsim}
\begin{figure}
\includegraphics[width=\columnwidth]{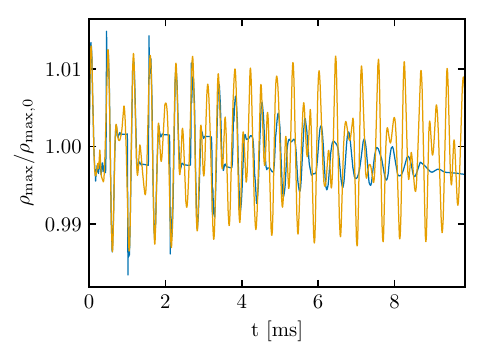}
\caption{Evolution of the maximum rest-mass density normalized to its initial value for a $1.4~M_\odot$ TOV neutron star modelled as a polytrope with $K=100$ and $\Gamma=2$. The blue line refers to a moving-mesh setup, while the orange line to a static-mesh setup. Both simulations adopt the Cowling approximation. In both cases, a radial velocity perturbation with amplitude $-0.005$ was applied. See the main text for details regarding the mesh setup.}
\label{fig:TOVCowrho}
\end{figure}

\begin{figure}
\includegraphics[width=\columnwidth]{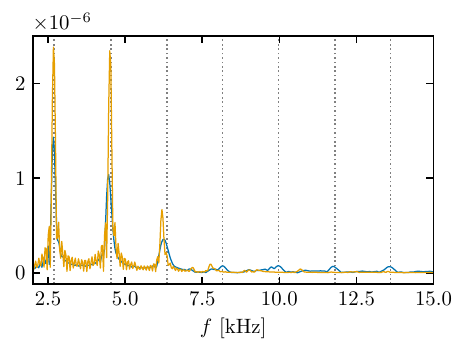}
\caption{Frequency spectrum of the maximum rest-mass density evolution of a $1.4~M_\odot$ TOV neutron star described by a polytropic EOS with $K=100$ and $\Gamma=2$ employing the Cowling approximation (see Fig.~\ref{fig:TOVCowrho}). The blue line corresponds to a moving-mesh simulation, while the orange line refers to a static-mesh calculation. The vertical dashed lines correspond to the frequencies computed in \citet{2002PhRvD..65h4024F}. The units of the vertical axis are arbitrary.}
\label{fig:TOVCowrhoFFT}
\end{figure}

Figure \ref{fig:TOVCowrho} shows the evolution of the maximum density normalized to its initial value. The blue line refers to a moving-mesh simulation, while the orange line to a static-mesh run where the mesh-generating points do not move. We note that both simulations preserve the initial TOV solution, i.e. the whole radial density profile, with only a minor density drift during the roughly $10$~ms evolution. The moving-mesh simulation features a stronger damping of the excited oscillations and a somewhat more pronounced density drift in comparison to the static-mesh case.  We extract the main radial pulsation frequency by a Fourier transform of the density oscillations. We obtain $2.672$~kHz for the moving mesh and $2.682$~kHz for the static mesh, which are both in excellent agreement with previous results \citep[$2.696$~kHz in][]{2002PhRvD..65h4024F}. In Fig.~\ref{fig:TOVCowrhoFFT} we present the power spectrum of the normalized maximum density (see Fig.~\ref{fig:TOVCowrho}). We consider the whole signal and apply a Tukey window with a shape parameter of $0.1$. In addition, we zero pad the signal, which effectively leads to smoother curves in the power spectrum. We note that a number of overtones are excited as well. In particular, in the moving-mesh simulation we identify $6$ overtones, which all agree within less than $2\%$ with the values reported in \citet{2002PhRvD..65h4024F}. The presence of several overtones is in line with the observation of several box-shaped oscillation cycles in~Fig.~\ref{fig:TOVCowrho} as the Fourier transform of a pulse wave is given by a number of higher overtones with decaying magnitude. In the static mesh simulation higher overtones seem to be less excited (or stronger damped) and only the first two appear prominently. We also extract the FWHM of the first three peaks in Fig.~\ref{fig:TOVCowrhoFFT} and find $129$~Hz, $130$~Hz and $132$~Hz ($161$~Hz, $212$~Hz and $321$~Hz) for the static-mesh (moving-mesh) calculations.

\begin{figure*}
\begin{center}
\resizebox{17cm}{!}{\includegraphics{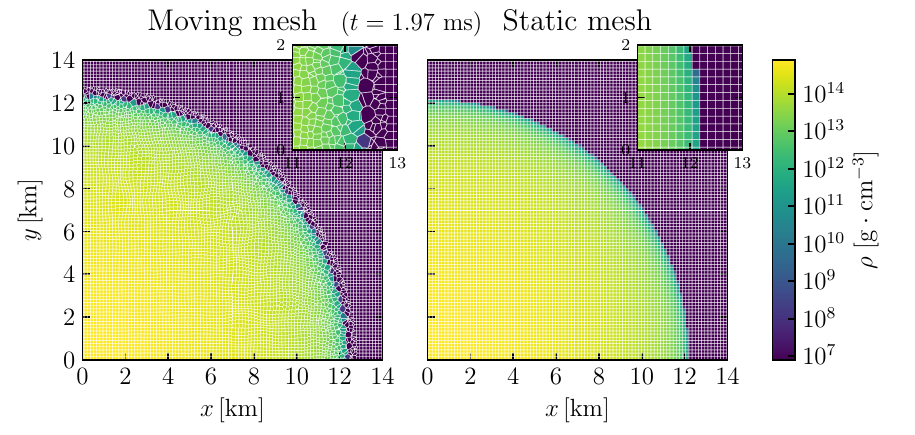}}\\
\caption{Rest-mass density at the $z=0$ plane for the moving-mesh (left panel) and static-mesh (right-panel) simulations on a fixed spacetime. The thin white lines represent the cell boundaries and thus display the mesh geometry. The subpanels at the top right corner of each panel depict a zoomed-in version of the region $[11,13]\times[0,2]$ in the respective plot. Both snapshots are taken at a time $t=1.97~\mathrm{ms}$.}
\label{fig:TOV_MT5_rho2_snap04_comp}
\end{center}
\end{figure*}

The moving-mesh and static-mesh evolutions are rather similar for the first few milliseconds. However, at later times the moving-mesh setup exhibits some damping in contrast to the static-mesh. This possibly originates to some extend from the surface layers. Initially the star contracts while atmosphere cells do not move. This results in a small gap between mesh-generating points which correspond to stellar material and thus closely follow the fluid motion, and points belonging to the atmosphere. This leads to larger cells close to the surface and thus the resolution at the stellar surface effectively drops. When the star expands, the stellar surface moves into the atmosphere. During expansion and contraction phases of the star, cells can cross the atmosphere threshold. Cells belonging to the star can become atmosphere, and atmosphere cells can accumulate material to become ``active'' stellar cells. Overall this leads to a decrease in the resolution close to the surface already after the first few ms. This behaviour is shown in Fig.~\ref{fig:TOV_MT5_rho2_snap04_comp}, which displays the rest-mass density in the $z=0$ plane after evolving the system for roughly $2$~ms. The left panel refers to the moving-mesh simulation, while the right panel corresponds to the static-mesh simulation. In both panels, we employ white lines to display cell boundaries, which reveals the mesh geometry. Figure~\ref{fig:TOV_MT5_rho2_snap04_comp} captures how the moving mesh evolves compared to the static-mesh simulation (i.e. also how the moving mesh evolves compared to the initial mesh geometry). In particular, the static-mesh and moving-mesh simulations have similar resolutions in the interior of the neutron star up to a few hundreds of meters beneath the surface. In the outer meters of the crust, the moving-mesh simulation has a lower resolution, which is one of the reasons for the higher damping. Finally, in the moving-mesh simulation a thin high-resolution shell forms right at the surface because of cells which originally belonged to either the star or the atmosphere.

\begin{figure}
     \begin{subfigure}[b]{\columnwidth}
         \centering
            \includegraphics[width=\textwidth]{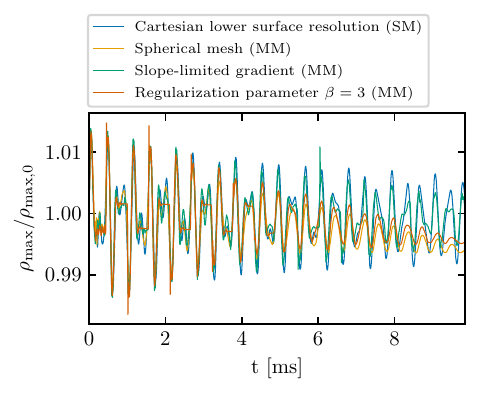}
            \caption{}\label{fig:TOV_MT4_AdditionalSims_rhonorm}
    \end{subfigure}

     \begin{subfigure}[b]{\columnwidth}
         \centering
            \includegraphics[width=\textwidth]{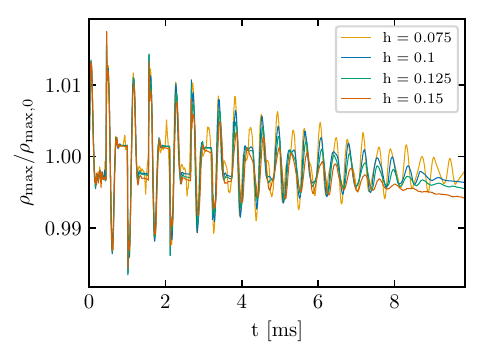}
            \caption{}\label{fig:TOV_MT4_ResStudy_rhonorm}
    \end{subfigure}
\caption{Rest-mass density evolution of a 1.4~$M_\odot$ TOV neutron star modelled as a polytrope with $K=100$ and $\Gamma=2$ within the Cowling approximation for different simulation setups. Panel (a): Effect of different numerical choices (see legend and main text in Sec.~\ref{subsec:TOVsim}). In the legend we denote moving- and static-mesh simulations as MM or SM, respectively. Panel (b): Impact of resolution on the moving-mesh calculations starting from a Cartesian initial mesh geometry. The default setup ($\mathrm{h}=0.1$) matches the blue line in Fig.~\ref{fig:TOVCowrho}.}
\label{fig:TOV_MT4_Add}
\end{figure}

To evaluate the effect that a lower resolution close to the surface has on the evolution, we perform an additional static-mesh simulation. We construct a mesh where we distribute the mesh-generating points to obtain a uniform Cartesian grid with a cell size of $0.1~M_\odot$ within a radius of $7.8~M_\odot\approx11.51~\mathrm{km}$ (i.e. $96\%$ of the stellar radius) surrounded by a uniform Cartesian grid with a cell size of $0.2~M_\odot$ outside the radius of $7.8~M_\odot$. We present the evolution of the rest-mass density in Fig.~\ref{fig:TOV_MT4_AdditionalSims_rhonorm}. We find that the evolution of the maximum density features some damping in this new static-mesh simulation and the amplitude of the oscillation has decreased by a factor of roughly $2$ after $\approx10~ms$ of evolution. In addition, we extract a frequency of $2.674$~kHz for the main radial pulsation, i.e. rather similar to the moving-mesh simulation. This observation supports that a lower resolution close to the surface can partially explain the damping in the maximum rest-mass density oscillation in the moving-mesh simulation.

Furthermore, we perform a set of additional simulations to investigate the oscillation behaviour in Fig.~\ref{fig:TOVCowrho} and we already refer to Sec.~\ref{subsec:TOVGR} showing that the use of the Cowling approximation is a major reason for damping in moving-mesh calculations. We investigate a number of aspects of the numerical description and their effects on the overall evolution:
\begin{enumerate}
\item \textit{Effect of resolution:} We perform three moving-mesh simulations, where the initial mesh-generating point distribution is similar to the one described in Sec.~\ref{subsec:TOVid}, but we vary the cell size of the central, high-resolution mesh to be $h=0.075$, $0.125$ and $0.15$ respectively (i.e. one higher-resolution and two lower-resolution simulations compared to the default setup). Figure~\ref{fig:TOV_MT4_ResStudy_rhonorm} shows the evolution of the maximum rest-mass density for the different resolutions. Increasing the resolution leads to less damping of the oscillation and reduces the minor density drift. The damping is not fully removed for the resolutions considered here. However, below we show that the setup for this resolution study (e.g. the initially Cartesian mesh) is not ideal and other aspects of the numerical treatment strongly improve the evolution. The extracted frequency of the main radial oscillation is $2.664$~kHz, $2.668$~kHz and $2.674$~kHz for $h=0.15$, $0.125$ and $0.075$ respectively, i.e. there is a very minor increase of the frequency with increasing resolution.

\item \textit{Mesh geometry:} We consider a moving-mesh simulation with an initial setup of mesh-generating points which leads to a spherical distribution of cells. The initial mesh is identical to the standard resolution setup described in detail in Sec.~\ref{subsec:TOVGRid}, where more details can be found. We emphasize that the equidistant radial separation between consecutive shells is $\approx191$~m, i.e. the resolution in this simulation is more comparable to the run with $h=0.125$ from point (i) above. We compare the number of cells within the star at the beginning of each simulation. The cells within the star in the simulation on the spherical mesh are $\approx92\%$ of the cells covering the star in the simulation on the (initially) Cartesian mesh with $h=0.125$. The differences in the number of cells covering the star between the spherical mesh simulation and the $h=0.075$, $0.1$ and $0.15$ simulations are larger. Hence, the spherical mesh simulation should be compared with the $h=0.125$ run in Fig.~\ref{fig:TOV_MT4_ResStudy_rhonorm}, while the first features a somewhat lower resolution. Evidently, the evolution of the rest-mass density on the spherical mesh features less damping compared to the (initially) Cartesian $h=0.125$ mesh (compare orange line in Fig.~\ref{fig:TOV_MT4_AdditionalSims_rhonorm} and green line in Fig.~\ref{fig:TOV_MT4_ResStudy_rhonorm} respectively). Using a spherical mesh does not completely remove the damping, but the comparison shows that a mesh which better captures the symmetries of the physical system is a more appropriate starting point for a moving-mesh simulation.

\item \textit{Gradient slope limiter:} We set up a moving-mesh simulation with an identical setup as the original one (i.e. Cartesian initial mesh with $h=0.1$), but we employ the standard slope-limited gradient in \textsc{Arepo} (Eq.~\eqref{ReconOrig}) instead of the MC slope limiter (see Sec.~\ref{subsec:Recon}). We find that the maximum density in this new simulation evolves in a rather similar manner to the original moving-mesh simulation with the MC slope limiter in the first few ms. However, at later times the maximum rest-mass density oscillation exhibits noticeably less damping in the new simulation and can still be clearly identified after roughly $10$~ms (see green line in Fig.~\ref{fig:TOV_MT4_AdditionalSims_rhonorm}). See also the discussion on slope limiters in \citet{2002PhRvD..65h4024F}. Employing the original slope-limiting procedure of \textsc{Arepo} may benefit from the fact that information about all the neighbouring cells enters the definition of $\psi_{ij}$ (see Eq.~\eqref{ReconOrig}) and not just the gradient estimate. Thus, it might be a more appropriate choice for moving-mesh simulations compared to the MC slope limiter.

\item \textit{Regularization scheme:} We consider the impact of the parameters $\beta$ and $f_\mathrm{shaping}$, which enter the regularization scheme\footnote{We note that in the standard setup these parameters are $\beta=2.25$ and $f_\mathrm{shaping}=0.5$ (see Sec.~\ref{subsec:MeshGeometry}).} (see Eq.~\eqref{Eq:RegVel} and Sec.~\ref{subsec:MeshGeometry}). We run a total of four additional moving-mesh simulations with an identical initial mesh as in the original run, where we systematically vary the values of the two parameters $(\beta, f_\mathrm{shaping})$ to be $(1.5, 0.5)$, $(3, 0.5)$, $(2.25, 0.3)$ and $(2.25, 0.7)$. Figure~\ref{fig:TOV_MT4_AdditionalSims_rhonorm} shows the impact of increasing $\beta$, i.e. applying less regularization as compared to the original simulation. The evolution features less damping than in the original setup and the oscillation can still be identified at the end of the simulation. We note that decreasing $\beta$ has the opposite effect, while changing the value of $f_\mathrm{shaping}$ has no noticeable effect (not shown). These results align with point (ii), namely that the moving-mesh approach performs better if less regularization of the mesh is required and thus the cell motion more closely follows the local fluid motion.

\end{enumerate}

\begin{figure}
     \begin{subfigure}[b]{\columnwidth}
         \centering
            \includegraphics[width=\textwidth]{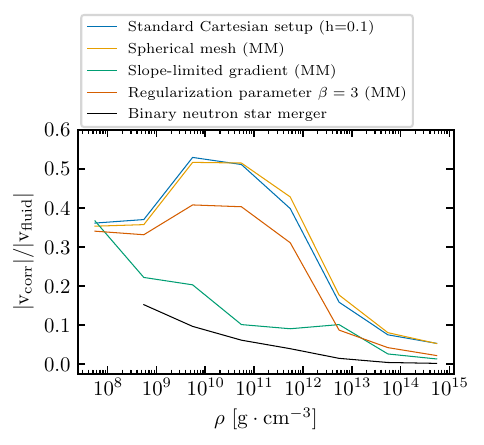}
            \caption{}\label{fig:TOV_MT4_AdditionalSims_CorVel}
    \end{subfigure}

     \begin{subfigure}[b]{\columnwidth}
         \centering
            \includegraphics[width=\textwidth]{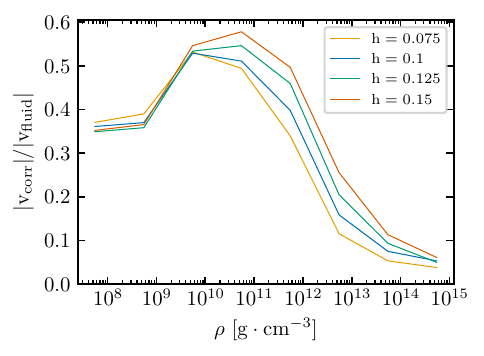}
            \caption{}\label{fig:TOV_MT4_ResStudy_CorVel}
    \end{subfigure}
\caption{Fraction of the corrective velocity due to regularization over the fluid velocity for the Cowling simulations of the TOV polytropic neutron star. The ratio is averaged over the density and over time (see main text). Panel (a): Includes all the moving-mesh setups from Fig.~\ref{fig:TOV_MT4_AdditionalSims_rhonorm}. The blue line refers to the standard (initially) Cartesian moving-mesh setup with $\mathrm{h}=0.1$ (note that the blue line in Fig.~\ref{fig:TOV_MT4_AdditionalSims_rhonorm} is a static-mesh run). The black line refers to the BNS merger simulation discussed in Sec.~\ref{sec:NSmergers}. Panel (b): Impact of resolution for all the (initially) Cartesian setups shown in Fig.~\ref{fig:TOV_MT4_ResStudy_rhonorm}. The blue line is the same as in panel (a).}
\label{fig:TOV_MT4_Add_CorVel}
\end{figure}

To quantify the impact of the regularization, we compute the fraction of cell motion stemming from the mesh regularization compared to the fluid velocity. Figure~\ref{fig:TOV_MT4_Add_CorVel} shows this fraction as a function of the density. We compute the fraction as an average over the density (by defining density intervals $[10^N,10^{N+1}]~\mathrm{g\cdot cm^{-3}}$ with $N$ being an integer in $[7,14]$) and over time by considering several snapshots (produced every $100$ code units, i.e.\ $\approx0.49$~ms). We report the different numerical setups in panel~\ref{fig:TOV_MT4_AdditionalSims_CorVel} and the results for different resolutions in panel~\ref{fig:TOV_MT4_ResStudy_CorVel}. We observe two general trends. First, the motion due to the mesh regularization is generally sizable (at all densities) and more pronounced in the outer layers, in line with the discussion above and Fig.~\ref{fig:TOV_MT5_rho2_snap04_comp}. Second, Figs.~\ref{fig:TOV_MT4_Add} and \ref{fig:TOV_MT4_Add_CorVel} clearly show that the magnitude of damping in moving-mesh simulations is related to the relative strength of mesh regularization. Numerical setups that require less regularization feature less numerical damping\footnote{As a reminder, the spherical mesh setup in panel~\ref{fig:TOV_MT4_AdditionalSims_CorVel} features a somehow lower resolution than the $\mathrm{h}=0.125$ setup in panel~\ref{fig:TOV_MT4_ResStudy_CorVel}. Namely, panel~\ref{fig:TOV_MT4_AdditionalSims_CorVel} shows that the fraction $|\bmath{v}_\mathrm{corr}|/|\bmath{v}_\mathrm{fluid}|$ for the (initially) spherical mesh is rather similar to the respective ratio for the (initially) Cartesian mesh which has a higher resolution.}. The effects discussed under points (i) to (iii) can thus be traced back to their impact on the mesh regularization. A more significant contribution from the mesh regularization to the cell motion arguably leads to more damping, as mesh regularization precisely means to introduce cell motion which does not follow the fluid motion and thus spoils the specific advantage of a moving mesh. For comparison, we also provide the relative fraction of the corrective cell velocity from regularization for the BNS merger simulation discussed in Sec.~\ref{sec:NSmergers}, which shows a much smaller relative impact of the mesh regularization.

Overall, we emphasize that the tests in this section demonstrate that further effort is required to identify setups which lead to better moving-mesh evolutions of TOV stars in the Cowling approximation. Based on our investigation, certain aspects of the numerical modeling (e.g.\ initial mesh geometry, slope limiter, regularization scheme parametrization) positively influence the moving-mesh evolution. We have chosen our default settings to be as close as possible to the setup in \citet{2002PhRvD..65h4024F} for a better comparison, but obviously other choices seem more appropriate for TOV systems. As apparent in Fig.~\ref{fig:TOV_MT4_Add_CorVel}, TOV stars may not be an easy target for moving-mesh simulations because of their quasi-stationary nature where, in contrast to dynamical systems like BNS mergers, the motion of the mesh is strongly determined by regularization.

We note that the cell rearrangement in the moving-mesh evolution highlights that a direct comparison between a static-mesh and a moving-mesh simulation is not necessarily straightforward. Even if the initial mesh geometries match, the moving mesh quickly rearranges and does not have a single resolution that one can compare to the fixed mesh. In addition, allowing the cells to move without taking the mass distribution of the system into account can lead to issues close to the surface as reported. The rearrangement of cells close to the surface can create small cells. If these cells are not derefined, which we do not do in our TOV simulations, they can in principle reduce the (global) timestep, hence increasing the required computational effort. We note that the mesh-generation algorithm typically requires more time to construct a Cartesian grid compared to other distributions with the same number of mesh-generating points due to the extra cost required to resolve geometric degeneracies during mesh construction.

\subsection{Dynamical spacetime}\label{subsec:TOVGR}

\subsubsection{Initial data}\label{subsec:TOVGRid}
We construct TOV data for a $1.41~M_\odot$ neutron star configuration (central density $\rho_c=9.545\times10^{-4}$) described by the H4 EOS \citep{2006PhRvD..73b4021L} modelled as a piecewise polytrope \citep{2009PhRvD..79l4032R}. We complement H4 with a $\Gamma_{\mathrm{th}} = 1.75$ thermal ideal-gas component.

The initial mesh-generating point distribution and subsequently the mesh geometry is different from our Cowling tests. We map the 1D TOV data to a spherical distribution of cells located at the center of the simulation domain. We use a total of $85$ shells extending up to a distance of $11~M_\odot\approx16.2$~km, with an equidistant radial separation $\approx191$~m between consecutive shells. In each shell we distribute $12N_\mathrm{side}^2$ cells based on the HEALPix tesselation by \citet{2005ApJ...622..759G}, where for a shell with inner radius $r_\mathrm{lower}$ and outer radius $r_\mathrm{upper}$ we set $N_\mathrm{side}=\sqrt{\pi/12} \times (r_\mathrm{lower}+r_\mathrm{upper})/(r_\mathrm{upper}-r_\mathrm{lower})$ \citep[see also][]{2012MNRAS.424.2222P}. In addition to these shells we place a coarse Cartesian grid to fill the rest of the computational domain. This setup is our standard resolution run. In addition, we construct two similar setups with $100$ shells (i.e. a resolution of $\approx162.4$~m) and $64$ shells (namely a resolution of $\approx253.7$~m), which we employ for higher and lower resolution (moving-mesh) simulations respectively. Finally, we also test a mesh which is similar to our standard resolution, but we randomly place the mesh-generating points on each shell to eliminate possible grid orientation effects.

In this section we compare the radial mode frequency from our simulations to a calculation with an independent linear perturbation code, which we developed following the approach outlined in \citet{1997A&A...325..217G}. Unlike Sec.~\ref{subsec:TOVCowling}, we do not compare to results from an independent Cartesian hydrodynamics code\footnote{We note that our perturbative code does not handle the Cowling approximation, which is why we cannot compare to perturbative results in Sec.~\ref{subsec:TOVCowling}.}. Since the perturbative result is practically exact and the comparison does not rely on choosing the same grid setup, we employ a grid that captures well the geometry of the physical system.

In contrast to the previous test, the spacetime evolves dynamically. We solve the metric field equations on a uniform Cartesian grid with $129^3$ points with a resolution $h_\mathrm{M}=0.3~M_\odot$. Similar to the Cowling tests, we excite the radial oscillation with a perturbation of the form \eqref{eq:RadPert} with $A=-0.001$ and we set $\rho_\mathrm{atm}=10^{-8}\times\rho_\mathrm{max}$ and $\rho_\mathrm{thr}=10\times\rho_\mathrm{atm}$.

\subsubsection{Simulations}\label{subsec:TOVGRsim}
\begin{figure}
\includegraphics[width=\columnwidth]{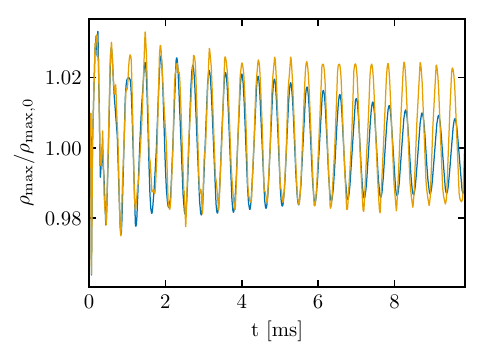}
\caption{Normalized maximum rest-mass density from a moving-mesh (blue line) and a static-mesh (orange line) evolution of a $1.41~M_\odot$ star described by the H4 EOS. The spacetime is evolved dynamically and the metric field equations are solved on a grid with $129^3$ points and resolution $0.3~M_\odot$. The pulsation is excited with a radial velocity perturbation with amplitude $-0.001$. See the main text for a description of the initial mesh geometry.}
\label{fig:TOVDynrho}
\end{figure}

In Fig.~\ref{fig:TOVDynrho} we present the time evolution of the normalized maximum density of the $1.41~M_\odot$ H4 stellar model with a moving mesh (blue) and a static mesh (orange) with our standard resolution setup. Again, we compute the fundamental radial pulsation frequency. Using a Fourier transform of the density oscillations we obtain $2.343$~kHz for the moving mesh and $2.358$~kHz for the static mesh. For comparison, the perturbative calculation gives $2.385$~kHz. Deviations of the order of one per cent are comparable to what is found by other codes e.g. \citep{2002PhRvD..65h4024F}.

\begin{figure}
\includegraphics[width=\columnwidth]{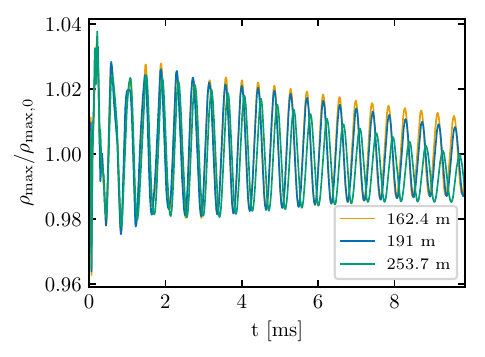}
\caption{Evolution of the normalized maximum rest-mass density in moving-mesh simulations with different resolutions. The legend provides the radial separation between consecutive shells in the initial (spherical) mesh. The spacetime is dynamically evolved. Note that the standard resolution setup (blue line), matches the blue line in Fig.~\ref{fig:TOVDynrho}.}
\label{fig:TOV_Dyn_Res}
\end{figure}

Moving-mesh runs with the higher and lower resolution (see Fig.~\ref{fig:TOV_Dyn_Res}) result in frequencies of $2.352$~kHz and $2.318$~kHz respectively, i.e. increasing the resolution leads to frequencies which lie closer to the perturbative result. In addition, we perform a moving-mesh simulation with an initial mesh setup with standard resolution ($\approx191$~m) including a random component to slightly offset the mesh-generating points. We obtain $2.349$~kHz, which is slightly higher than the result from the same resolution setup without the random component. We note that including the random component in the mesh setup reduces grid effects, while it only slightly increases the damping in the maximum density oscillation. The overall agreement in the frequencies validates our implementation of GRHD and the metric solver, as well as their coupling in a realistic setup which employs a microphysical EOS.

The moving-mesh and static-mesh standard resolution simulations of the star preserve the initial TOV solution during the whole simulation time with only a very mild drift in the central density, which is also seen in other simulations \citep[e.g.][]{2002PhRvD..65h4024F}. The drift diminishes with higher resolution, see Fig.~\ref{fig:TOV_Dyn_Res}. Increasing the resolution also yields less damping in the rest-mass density oscillation.

The oscillations in the moving-mesh simulations show some damping over time, but the amplitude is still sizable at the end of the calculations. In comparison to the Cowling runs, the damping is less pronounced, and the evolution of $\rho_\mathrm{max}$ generally does not deviate as much from the respective static-mesh runs. In Sec.~\ref{subsec:TOVsim} we found that a combination of effects enhances damping in moving-mesh simulations compared to static-mesh runs in the Cowling approximation (e.g. mesh geometry, slope limiter, regularization). The damping of the moving-mesh simulation in Fig.~\ref{fig:TOVDynrho} (dynamical space time) in comparison to the different moving-mesh runs in Sec.~\ref{subsec:TOVsim} (Cowling) is relatively moderate. This suggests that the Cowling approximation (versus a dynamical evolution of the space time) is another major reason for the damping observed in moving-mesh simulations in Sec.~\ref{subsec:TOVsim}. Although we compare simulations with different setups (EOS, initial excitation, mesh geometry), this is in line with the observation that the response to perturbations is different in Cowling and dynamical space time simulations \citep[see e.g.][]{2006MNRAS.368.1609D}. Apart from the aspects discussed in Sec.~\ref{subsec:TOVsim}, specific implementations such as modifying the cell motion close to the surface and suppressing small secular drifts of cells are likely to improve the behavior in quasi-stationary situations. Since our work is targeted to highly dynamical problems, where we do not face the same issues, we do not follow up on these points here.


\section{Binary neutron star mergers}\label{sec:NSmergers}
In this section we discuss a BNS merger simulation. This is the first simulation of a neutron star merger using a moving mesh and we show that this approach can be successfully used to simulate such systems.

\subsection{Initial data and setup}\label{subsec:BNSid}
We employ the DD2 EOS \citep{Hempel2010,Typel2010} as a zero-temperature $\beta$-equilibrium tabulated microphysical EOS. We remark that the DD2 model provides the full temperature and composition dependence of the EOS. In this work however, for convenience, we only use a slice at $T=0.1$~MeV as the lowest temperature provided by the EOS table. We supplement the barotropic EOS with a thermal ideal-gas component with $\Gamma_\mathrm{th}=1.75$, as described in Sec.~\ref{subsec:EoS} \citep[see][for more details]{2010PhRvD..82h4043B}. DD2 is marginally compatible with current observational constraints on the tidal deformability from GW170817 \citep{2017PhRvL.119p1101A,2019PhRvX...9a1001A} and fully consistent with mass measurements of various binary systems \citep{2010Natur.467.1081D,2013Sci...340..448A,2018ApJS..235...37A,2018ApJ...859...54L,2020NatAs...4...72C}.

We construct initial data for an equal-mass BNS system in a quasi-circular quasi-equilibrium orbit using LORENE\footnote{\url{http://www.lorene.obspm.fr/}} \citep{2001PhRvD..63f4029G}. The two companion neutron stars have a gravitational mass of $M=1.35~M_\odot$ (at infinite binary separation) and are irrotational. The initial separation is $26~M_\odot\approx38.4$~km. LORENE solves the metric field equations using the conformal flatness approximation like our code. As a result, we do not observe an unphysical transient at the beginning of our simulation as compared to fully relativistic simulations, which react to the missing GW content of the initial data. Hence, we can start our simulation from a relatively small initial separation of the two companion stars. 

We map the initial data from LORENE to a distribution of mesh-generating points which follows approximately the mass distribution. In particular, around the centers of each star, we construct spherical shells and then distribute cells on each shell based on the HEALPix algorithm by \citet{2005ApJ...622..759G} to obtain a distribution of mesh generating points which have roughly the same distance within the shell. We impose that the distance between points within the shell equals the thickness of the shell. The grid setup is described in detail in \citet{2017A&A...599A...5O}. The original method described in \citet{2017A&A...599A...5O} determines the radial positions of the shells such that each cell has roughly a mass $m_\mathrm{cell,0}$ following the density profile of the initial data. Here $m_\mathrm{cell,0}$ is a free parameter, which directly relates to the resolution. The procedure described in \citet{2017A&A...599A...5O} assumes that the density profile of each star is spherically symmetric. Hence, in our case, we employ a TOV solution with $M=1.35~M_\odot$ modelled by DD2. We emphasize that the TOV solution is only relevant for the purpose of distributing the mesh-generating points in a way that closely follows the mass distribution within the binary (around each companion star). The actual initial data, e.g. the three-dimensional density profile, originate from LORENE.

In the case of neutron stars, this procedure would lead to large cells close to the stellar surface, where the density drops steeply. Therefore, we modify the employed TOV density profile in the outer regions of the star. Between the distances $r_\mathrm{in}=r(\rho_c/2)$ and $r_\mathrm{out}=1.1\times R$, we keep $\rho\sqrt{\gamma}=\rho\psi^6$ fixed to the value at $r_\mathrm{in}$. Here $\rho_c$ is the central density and $R$ the radius of the TOV star.  We determine the position of the radial shells based on this modified density profile, which enhances the resolution in the outer layers of the star compared to the original procedure. As stressed, we finally assign the values from the LORENE solution to the cells which we obtain through this procedure.

This grid extends beyond the stellar radius. This particular setup guarantees that the resolution does not drop abruptly close to the surface because of the steep density gradient. Moreover, we resolve the regions around $R$, which is important because the stars in the binary are not perfect spheres, but deformed. In the current simulation, we set $m_\mathrm{cell,0}=1.68\times10^{-6}~M_\odot$. In principle this construction leads to a mesh with no grid orientation effects \citep[see][for more details]{2017A&A...599A...5O}. We cover (vacuum) regions outside the spheres with radius $r_\mathrm{out}$ around each star with a coarse uniform Cartesian mesh with resolution $10.1~M_\odot$. The atmosphere density is set to $\rho_\mathrm{atm}=10^{-7}\times\rho_\mathrm{max}$ and $\rho_\mathrm{thr}=10\times\rho_\mathrm{atm}$. We do not impose any symmetries during the simulation. The metric field equations are solved on a uniform Cartesian grid with $129^3$ points and resolution $0.8~M_\odot$.

During the simulation we allow for cell refinement and derefinement. We set a cell target mass $m_\mathrm{cell,0}=1.68\times10^{-6}~M_\odot$, which is the same that we used to find the radial positions of cells in the mesh-constructing algorithm \citep[see][for more details]{2017A&A...599A...5O}. We refine cells with mass $m_\mathrm{cell}>2\times m_\mathrm{cell,0}$ and derefine cells with mass $m_\mathrm{cell}<m_\mathrm{cell,0}/2$. Furthermore, for each cell we check the volume of every neighbouring cell. We ensure that a cell is not derefined if $V>1.5\times V_\mathrm{ngb}^\mathrm{min}$, where $V$ the cell volume and $V_\mathrm{ngb}^\mathrm{min}$ the volume of the smallest neighbouring cell. We refine any cell for which $V>5\times V_\mathrm{ngb}^\mathrm{min}$ holds. To avoid creating an irregular mesh through the refinement process, in all cases we do not refine highly distorted cells i.e.\ cells with $\alpha_\mathrm{max}\geq3.375$ \citep[see][for more details]{2012MNRAS.425.3024V}. Cell refinement takes place if any of the refinement criteria is satisfied, while to derefine a cell we require that all derefinement criteria allow derefinement at the same time. This combination of criteria guarantees that we have a mesh where many cells have comparable mass content, while we also resolve the surface with a decent resolution (i.e.\ during the first milliseconds cells in the crust have an average size of $\approx0.3~M_\odot$, which is larger than typical cell sizes in current static-mesh calculation; in the neutron star cores cells have typical dimensions of about $0.1~M_\odot$ which is comparable with typical cell sizes in static-mesh simulations). Roughly $1.7\times10^6$ cells with $\rho>\rho_\mathrm{thr}$ resolve physically interesting regions. Notably, shortly after the beginning of the simulation, we find only a small number of cells outside the two stars due to cell derefinement.  Hence, in our approach the hydrodynamical domain can be arbitrarily large with practically no effect on the computational time.

We compare the maximum density (see Fig.~\ref{fig:BNSrhomax}) and lapse function evolution during the first few milliseconds of our binary system evolution to an independent simulation of a single $1.35~M_\odot$ TOV star described by DD2. The mesh-generating point setup in the isolated star simulation is identical to the one which we employ for each individual star in the binary, while we keep the same metric resolution, atmosphere parameters and refinement/derefinement criteria. Discretization errors by the finite resolution excite oscillations in the maximum density and the minimum lapse function. Comparing the calculations of the isolated star and the binary, the frequencies are similar, while the amplitudes are slightly higher in the case of the isolated neutron star. Hence, we conclude that the setup of our binary initial data works robustly and that the procedure does not introduce additional errors apart from those which are expected, i.e. truncation errors.

We also perform additional simulations where we vary certain aspects of the numerical treatment (e.g.\ the hydro and metric resolutions) to evaluate their impact on the results. The additional simulations are discussed in Appendices~\ref{subapp:NumSetupRes} and \ref{subapp:NumSetupFreq}.

\subsection{Simulations}\label{subsec:BNSsim}
\subsubsection{General dynamics}\label{subsec:BNSdyn}
\begin{figure}
\includegraphics[width=\columnwidth]{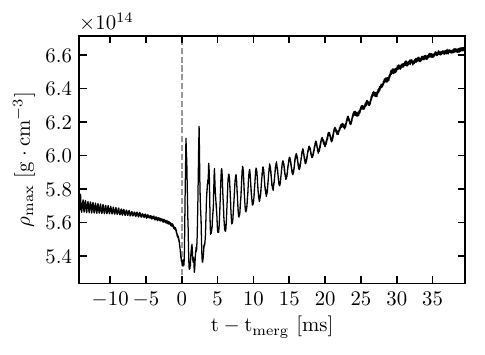}
\caption{Maximum rest-mass density as a function of time for a BNS system with two $1.35~M_\odot$ neutron stars modelled with the DD2 EOS. The vertical dashed line indicates the time of merging.}
\label{fig:BNSrhomax}
\end{figure}

\begin{figure*}
\begin{center}
\resizebox{17cm}{!}{\includegraphics{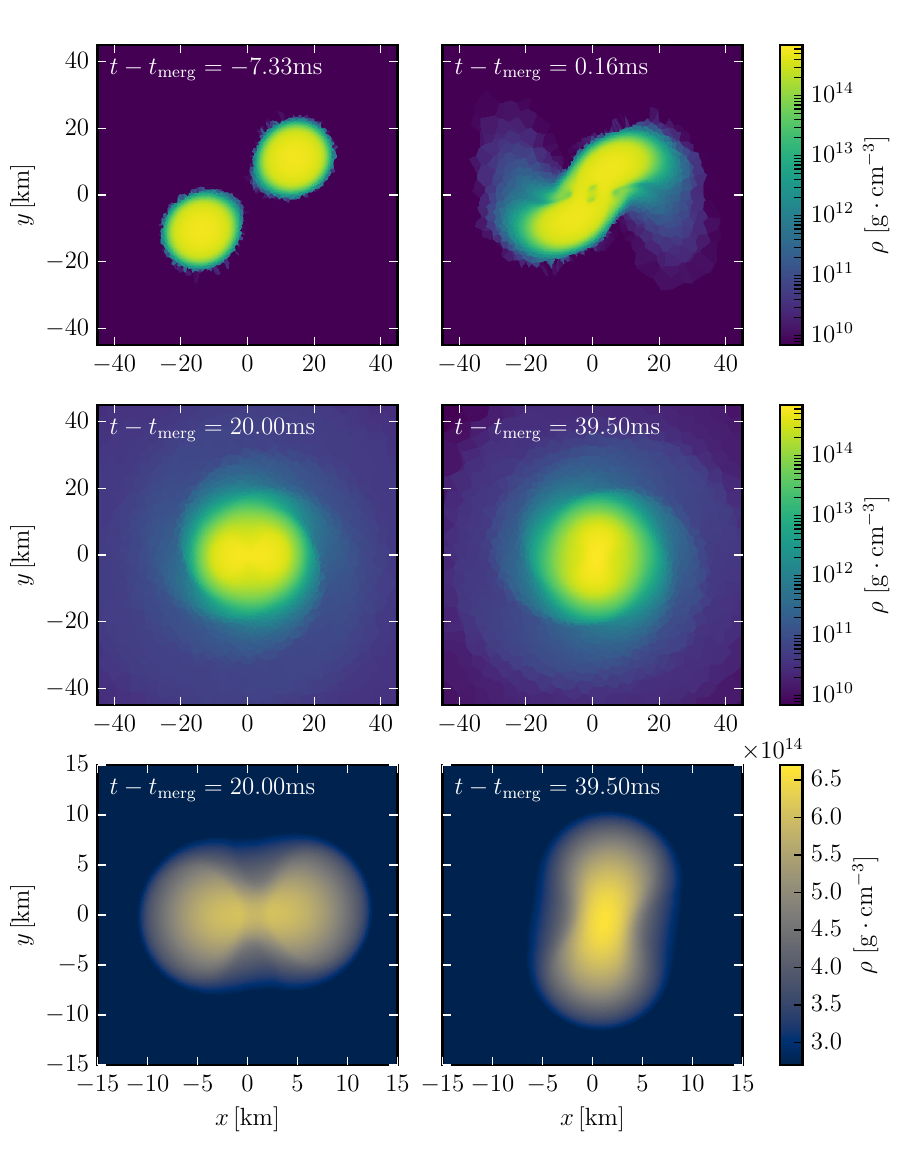}}\\
\caption{Evolution of the rest-mass density for the BNS merger simulation. Each panel shows a slice through the orbital plane of the binary. The densities in the upper and middle rows are displayed on a logarithmic scale, while the bottom row focuses on the high-density material of the remnant employing a linear density scale. The times are chosen such that the top row panels show snapshots from the late inspiral stage and the moment of merging, while the middle row panels display very late stages of the post-merger evolution. The times in the bottom row panels match those in the middle row.}
\label{fig:BNS_rho}
\end{center}
\end{figure*}

Figure \ref{fig:BNSrhomax} shows the evolution of the maximum rest-mass density $\rho_\mathrm{max}$ during the simulation. The vertical dashed line indicates the time when the two neutron stars merge $t_\mathrm{merg}$\footnote{We define $t_\mathrm{merg}$ as the time when the GW signal amplitude reaches its maximum.} and separates the inspiral and the post-merger phase. The main stages of the binary evolution are shown in Fig.~\ref{fig:BNS_rho}, where we present the rest-mass density in the orbital plane of the binary at four different times. The top left and top right panels correspond to snapshots taken during the late inspiral\footnote{Althouh we simulate only the very last phase of the inspiral a few orbits before merging, we will throughout this paper refer to this part of the simulation as ``inspiral'' for brevity.} and right after merging, respectively. Middle and bottom row panels present two times at the late stages of the post-merger evolution on a logarithmic and on a linear scale. We evolve the binary for $\approx39.5$~ms after the stars merge.

Throughout the inspiral the two neutron stars revolve around each other, while the orbital separation decreases due to energy and angular momentum losses by GWs. As mentioned, during the inspiral the maximum density features small oscillations (Fig.~\ref{fig:BNSrhomax}) because discretization errors excite dominantly the radial mode. As the binary components approach each other, tidal effects become more pronounced \cite[an incomplete list of early hydrodynamical studies of the merger include e.g.][]{1994ApJ...432..242R,1994PhRvD..50.6247Z,1996PhRvD..54.1317W,1997A&A...321..991R,1999A&A...341..499R,2002PhRvD..65j3005O,2005PhRvD..71h4021S,2008PhRvD..78h4033B}. The tidal deformations are visible in the top left panel of Fig.~\ref{fig:BNS_rho}, which corresponds to the last few revolutions before the stars merge.

\begin{figure*}
\begin{center}
\resizebox{17cm}{!}{\includegraphics{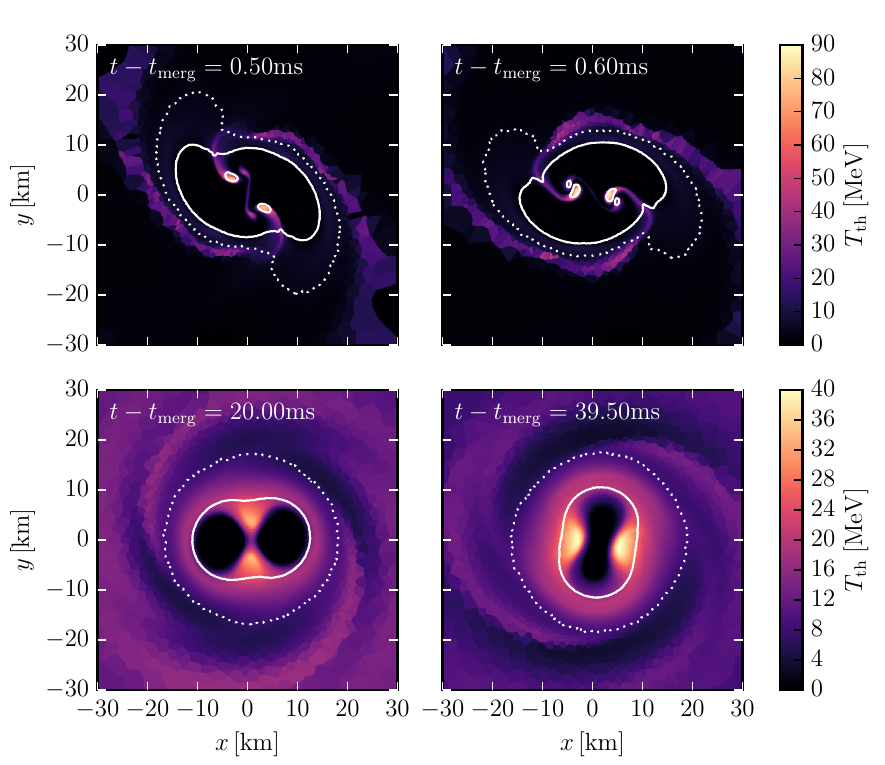}}\\
\caption{Temperature in the orbital plane for the BNS merger simulation. The top row shows snapshots right after the neutron stars merge, when the temperature reaches the highest value of $T_{\mathrm{th}}\approx90$~MeV. The bottom row presents snapshots taken at $20$~ms and $39.5$~ms after merging. White dotted and solid contours indicate densities of $10^{13}~\mathrm{g \cdot cm^{-3}}$ and $2.7\times10^{14}~\mathrm{g \cdot cm^{-3}}$, respectively. The two cold, high-density cores are clearly visible at late stages of the evolution (bottom panels).}
  \label{fig:BNS_temp}
\end{center}
\end{figure*}

The stars collide with a relatively large impact parameter. The collision results in the sudden increase of the maximum density immediately after $t_\mathrm{merg}$ in Fig.~\ref{fig:BNSrhomax} and shock-heating of material at the collision interface \citep[e.g.][]{1996A&A...311..532R,1999A&A...341..499R,2005PhRvD..71h4021S,2007A&A...467..395O}. Figure \ref{fig:BNS_temp} shows snapshots of the temperature $T_{\mathrm{th}}$ in the orbital plane right after merging (top row), when the system reaches the highest temperatures, as well as two snapshots in the post-merger phase (bottom row). As shown in the upper panels of Fig.~\ref{fig:BNS_temp}, matter at the collision interface of the two neutrons stars reaches temperatures of almost $90$~MeV. In Fig.~\ref{fig:BNS_temp} we also overplot contour lines corresponding to two different densities. The white dashed line indicates where the density equals $10^{13}~\mathrm{g \cdot cm^{-3}}$ to highlight the region containing the bulk of matter at the times shown in the plots. The solid white line corresponds to a density of $2.7\times10^{14}~\mathrm{g \cdot cm^{-3}}$ (nuclear saturation density). Most of the high-density parts of the stars, except for matter at the collision interface, remain cold even during the merger phase \citep[see e.g.][]{2005PhRvD..71h4021S,2007A&A...467..395O,2016PhRvD..94d4060K,2017PhRvD..96d3004H}. The highest temperatures are reached in hotspots with densities slightly below $2.7\times10^{14}~\mathrm{g \cdot cm^{-3}}$. The densities are lower in these blobs because of the significant thermal pressure at these temperatures. These regions are also visible in the top right panel of Fig.~\ref{fig:BNS_rho} and form due to the mixing of material from the two stars \citep[see e.g.][]{2016PhRvD..94d4060K}. The contact interface between the two stars is subject to the Kelvin-Helmholtz instability. The distribution and evolution of the temperature in the upper row of Fig.~\ref{fig:BNS_temp} is indicative of the local vorticity \citep[see e.g.][]{2015PhRvD..92l4034K}.

In the post-merger phase, a double-core structure forms. We observe that our simulation preserves the double-core structure for more than $20$~ms after $t_\mathrm{merg}$. This is clearly shown in the middle and bottom rows of Fig.~\ref{fig:BNS_rho}, as well as the bottom row of Fig.~\ref{fig:BNS_temp}. The two cores (enclosed in the white solid contour line in the bottom left panel of Fig.~\ref{fig:BNS_temp}) can be clearly identified $20$~ms after merging. The centers of the cores merge at $t-t_\mathrm{merg}\approx28$~ms. Even at the very late stages of the evolution, i.e. at $t-t_\mathrm{merg}=39.5$~ms, the high-density material has not yet settled to a single spherically-shaped core, but exhibits a bar-shaped structure. The double-core phase in our simulation lasts significantly longer compared to other simulations with fixed-grid finite-volume approaches \citep[see e.g.][]{2016PhRvD..94d4060K,2017PhRvD..96d3004H}. Similarly, the quasi-radial mode survives for a long time after merging as shown in Fig.~\ref{fig:BNSrhomax}.

The cores remain cold during the whole post-merger evolution. The remnant still exhibits high temperatures, up to $\approx40$~MeV. These highest temperatures at late times occur at the outer edges of the shearing interface between the two cold cores. The system exhibits density spiral arms starting at the central object during the whole post-merger phase that we simulate, as can be seen in both middle row panels of Fig.~\ref{fig:BNS_rho} and in the bottom row panels of Fig.~\ref{fig:BNS_temp}.

We cannot identify the development of a pronounced $m=1$ instability in the evolution of the rest-mass density in the orbital plane until the end of the simulation \citep{2015PhRvD..92l1502P,2016PhRvD..94d3003L,2016PhRvD..93b4011E,2016PhRvD..94f4011R}. Based on a modal decomposition of the density \citep[see e.g.][]{2016PhRvD..94f4011R}, the $m=2$ mode features only extremely minor damping and remains dominant compared to the $m=1$ mode throughout the whole simulated post-merger phase. For comparison, we simulate the binary system with an SPH code, which also adopts the conformal flatness approximation \citep{2002PhRvD..65j3005O,2007A&A...467..395O}. We employ the same EOS treatment and choose a resolution of roughly $3\times10^5$ SPH particles in total\footnote{The SPH particle number as measure of the resolution should not be directly compared to the number of resolving elements in a grid code. Effectively, the resolution of the SPH run is lower compared to the moving-mesh simulation.}. In the SPH simulation the $m=1$ instability does clearly occur for the same binary system. A modal decomposition of the SPH simulation reveals that the $m=2$ mode is damped over time and at $t-t_\mathrm{merg}\approx15$~ms its amplitude decays below that of the $m=1$ mode. Considering that the two codes employ very similar metric modules, these results support that the development of the one-arm instability depends on the employed hydrodynamics schemes and the resolution.

\begin{figure}
\includegraphics[width=\columnwidth]{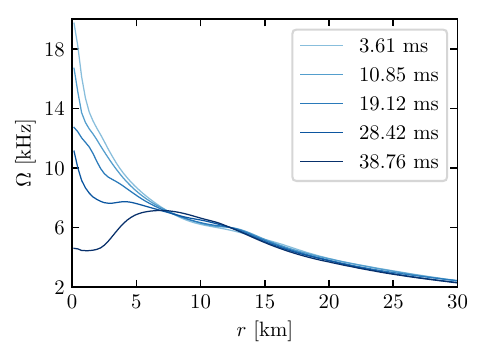}
\caption{Rotation profile of the remnant at different times after merging. The angular velocity $\Omega$ is averaged along the azimuthial direction and over a time interval of $1$~ms. The legend indicates $t-t_\mathrm{merg}$ for each line, where $t$ refers to the midpoint of the respective $1$~ms time interval.}
\label{fig:BNSOmega}
\end{figure}

We briefly examine the angular velocity profile\footnote{We define the angular velocity as $\frac{xv^y-yv^x}{x^2+y^2}$ where $v^i=u^i/u^0$. All quantities are computed with respect to the center of mass of matter with $\rho>0.95\times\rho_\mathrm{max}$ and we consider time- and azimuthially-averaged profiles.} in the equatorial plane at different stages of the post-merger evolution in Fig.~\ref{fig:BNSOmega}. The rotation profile initially exhibits a maximum at the center of the remnant. The value of the time- and azimuthially-averaged angular velocity $\Omega$ at the center decreases over time and at later times an off-center peak forms. The off-center maximum first appears at $t-t_\mathrm{merg}\approx28$~ms at a radial distance of $\approx4$~km. Over time the position of the peak moves outwards from the center to about $\approx7$~km at the end of the simulation. The qualitative characteristics of the angular velocity profile agree with what is reported in other simulations \citep[see e.g.][]{2005PhRvD..71h4021S,2016PhRvD..94d4060K,2017PhRvD..96d3004H,2017MNRAS.471.1879G}, but the evolution of the profile and the overall angular momentum redistribution happens over longer timescales. A similar delay in the evolution of the $\Omega$ profile has been reported in \citet{2018PhRvD..97l4039K} for simulations with very high resolution. The latter calculations, however, employ a different EOS and include magnetic fields, which is why a direct comparison is difficult.

Overall, the general dynamics and the qualitative features of our simulation are consistent with what is found in other simulations. A notable difference is that the double-core structure and the quasi-radial oscillations persist for a longer time. Similarly, an off-center peak of the angular velocity profile emerges only at relatively late times. These points hint that the evolution with the moving-mesh setup might have low numerical viscosity (see also the discussion on the damping of the GW signal in Sec.~\ref{subsec:BNSgw}).

Finally, we comment on the resolution of our simulation. In principle, we cannot define a single resolution in moving-mesh simulations, because cells do not have a fixed shape and volume. This is clearly visible in both Figs.~\ref{fig:BNS_rho} and \ref{fig:BNS_temp}, where in lower density regions the cell shapes are visible. We can however estimate the resolution under some assumption for the cell geometry. Here we assume that cells are spheres and focus on the high-density regions. For material with $\rho > 0.5\times\rho_\mathrm{max}$ at a given time we compute the average cell volume and in turn the cell radius. We then estimate the mean distance between cell centers in these regions as twice the computed radius. Throughout the post-merger phase, we obtain an average distance between cell centers of approximately $0.11~M_\odot\approx162$~m in regions with rest-mass density above $50\%$ of $\rho_\mathrm{max}$. Naturally, some cells are smaller (or larger) than what this number indicates. This highlights the ability of our implementation to reach resolutions which are roughly comparable to what is currently used in merger simulations, although typically with high-order schemes. Our simulation required a few weeks of computing time running on $192$ cores (see also Appendix~\ref{subapp:NumSetupRes}). In the future we plan to perform simulations with even higher resolution.

\subsubsection{Gravitational wave signal}\label{subsec:BNSgw}
\begin{figure}
\includegraphics[width=\columnwidth]{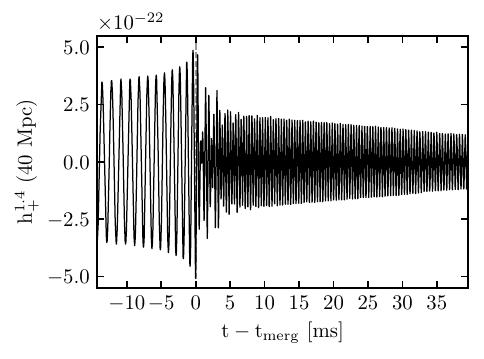}
\caption{Gravitational wave amplitude $h_+^\mathrm{1.4} = 1.4 \times h_+$, where $h_+$ is the strain of the plus polarization at a distance of $40$~Mpc for the BNS simulation. The vertical dashed line shows the merging time.}
\label{fig:BNSGWsignal}
\end{figure}

In Fig.~\ref{fig:BNSGWsignal} we show the plus polarization of the GW signal $h_+$ (strain at a distance of $40$~Mpc along the polar axis), multiplied by a factor $1.4$ (denoted as $h_+^\mathrm{1.4}$) to account for the underestimation of the amplitude by the quadrupole formula \citep[see][likely the factor is closer to 2 comparing recent simulations \citep{2022PhRvD.105d3020S,2022EPJA...58...74D}]{2005PhRvD..71h4021S}. The vertical dashed line in Fig.~\ref{fig:BNSGWsignal} indicates the merging time. Before $t_\mathrm{merg}$ the system emits GWs with a frequency twice as large as the orbital frequency. As the stars approach each other, the frequency, as well as the amplitude of the GW signal, increase. The coalescence of the stars excites a number of modes in the post-merger remnant, which shape the post-merger GW signal. 

Notably, the damping of the post-merger GW signal is very slow for the approximately $39.5$~ms of post-merger evolution, which is in agreement with our observation in Sec.~\ref{subsec:BNSdyn} that numerical viscosity may be relatively low in this simulation (see also Appendix~\ref{subapp:NumSetupRes}). We determine the damping time to be $\tau_\mathrm{peak}\approx48$~ms based on the analytic model presented by \citet{2022PhRvD.105d3020S}. The SPH simulation with roughly $3\times10^5$ SPH particles (see Sec.~\ref{subsec:BNSdyn}) yields $\tau_\mathrm{peak}\approx10.5$~ms. \citet{2022PhRvD.105d3020S} perform fully general relativistic simulations with finest grid resolutions of $277$~m and $185$~m on a fixed grid and obtain $\tau_\mathrm{peak}<11$~ms for all their models and resolutions ($\tau_\mathrm{peak}\approx7$~ms for the binary system with total gravitational mass of $2.7~M_\odot$), which is significantly below our current result. We do however note that \citet{2022PhRvD.105d3020S} employ a different EOS, which does not allow for a direct comparison with our simulation. We note that the GW backreaction scheme somewhat underestimates the physical damping by GWs because of the underestimated amplitude. In Appendix~\ref{subapp:NumSetupRes} we clarify that accounting for this underestimation does somewhat reduce the damping time scale, but we still find a slowly damped post-merger GW emission.

\begin{figure}
\includegraphics[width=\columnwidth]{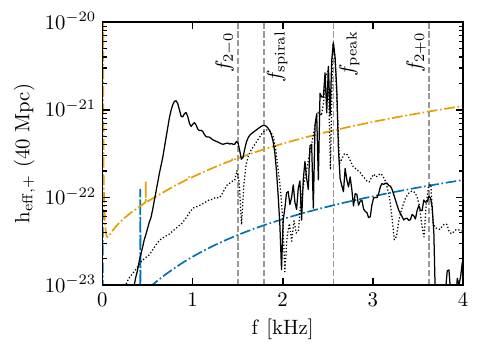}
\caption{Gravitational wave spectrum of the plus polarization at a distance of $40$~Mpc. The solid line refers to the whole simulation, while the dotted line displays the spectrum of only the post-merger phase. Both lines refer to $h_+^\mathrm{1.4} = 1.4 \times h_+$, as presented in Fig.~\ref{fig:BNSGWsignal}. The vertical dashed lines indicate the frequency peaks $f_\mathrm{peak}$, $f_\mathrm{spiral}$, $f_\mathrm{2-0}$ and $f_\mathrm{2+0}$. The upper (orange) and lower (blue) dash-dotted lines denote the design sensitivity of Advanced LIGO \citep{2015CQGra..32g4001L} and the Einstein Telescope \citep{2010CQGra..27s4002P}, respectively.}
\label{fig:BNSGWsignalFourier}
\end{figure}

In Fig.~\ref{fig:BNSGWsignalFourier} we present $h_\mathrm{eff,+}(f)=f\cdot\tilde{h}_\mathrm{+}(f)$, where $\tilde{h}_\mathrm{+}(f)$ is the Fourier transform of the $h_+^\mathrm{1.4}$ polarization, as measured at $40$~Mpc. The solid line corresponds to the spectrum of the full GW signal from the whole evolution, while the dotted line is the Fourier transform of the signal from the post-merger phase alone. In addition, we show the design sensitivity of Advanced LIGO \citep{2015CQGra..32g4001L} and the Einstein Telescope \citep{2010CQGra..27s4002P} with the upper and lower dash-dotted lines, respectively. We extract the main oscillation frequency in the spectrum (marked by a vertical dashed line) at $f_\mathrm{peak}=2.56$~kHz. For comparison, in our SPH simulation of the same binary system (see Sec.~\ref{subsec:BNSdyn}) we obtain $f_\mathrm{peak}=2.62$~kHz, which is in good agreement with our current result.  In comparison to the SPH simulation, the features in the GW spectrum in Fig.~\ref{fig:BNSGWsignalFourier} are more pronounced likely due to the low numerical damping of the signal. In addition, we compare to fully general relativistic simulations of the binary system from the CORE database\footnote{\url{http://www.computational-relativity.org}} \citep{2018CQGra..35xLT01D}. We extract $f_\mathrm{peak}=2.57$~kHz and $f_\mathrm{peak}=2.65$~kHz for the two available simulations with finest grid resolutions of $0.125~M_\odot$ and $0.083~M_\odot$ respectively \citep[see][for more details]{2016MNRAS.460.3255R,2017ApJ...842L..10R}. Both frequencies agree rather well with our result noting that these static-mesh simulations employed the full temperature-dependent EOS table.

The spectrum in Fig.~\ref{fig:BNSGWsignalFourier} contains several subdominant features, which are in principle observable. In particular there is a pronounced broad peak at about $1.8$~kHz. This peak emerges from the rotation of two antipodal tidal bulges, which form right after the merging. These tails rotate at a slower rate than the high-density parts and produce a peak at a frequency $f_\mathrm{spiral}$ \citep{2015PhRvD..91l4056B}. In this particular simulation we obtain $f_\mathrm{spiral}=1.79$~kHz, roughly $200$~Hz lower than in the SPH simulation. The amplitude of the peak in the moving-mesh simulation is in comparison to the SPH calculation increased. Since the gravity solver is identical in both simulations, the different properties of this secondary peak hint to some sensitivity of this feature to the hydrodynamics scheme.

In Fig.~\ref{fig:BNSGWsignalFourier} we also indicate the frequencies $f_\mathrm{2\pm0}$ originating from the non-linear coupling between the dominant oscillation mode $f_\mathrm{peak}$ and the quasi-radial oscillation mode $f_\mathrm{0}$ \citep{2011MNRAS.418..427S}. To identify this feature, we estimate  $f_\mathrm{0}$ from the evolution of the lapse function and tag the peaks which occur at $\approx f_\mathrm{peak}\pm f_\mathrm{0}$ in the GW spectrum.


\section{Summary}\label{sec:Conclusions}

In this work we extend the (originally Newtonian) moving-mesh \textsc{Arepo} code to general relativistic hydrodynamics employing the flux-conservative Valencia formulation. We couple the implementation with a solver of the Einstein field equations imposing the conformal flatness approximation. This new tool can in principle be applied to a variety of astrophysical scenarios including those that require the dynamical evolution of the spacetime. In this work we focus on applications to neutron stars and neutron star mergers and supplement the code with a module to include a high-density EOS.

We validate the implementation by performing different test calculations, which can be compared to independent results. We simulate isolated, static neutron stars with a fixed spacetime (Cowling approximation) and with a dynamical spacetime (TOV tests). In both tests the code preserves very well the density profile of the initial equilibrium model. The frequencies of radial oscillations, which are excited by truncation errors and an added perturbation, coincide very well with perturbative results and simulations from other codes, which verifies our implementation. We run simulations with moving meshes and static meshes which allow a more direct comparison to existing calculations. \textsc{Arepo} offers the advantage that the initial grid configuration can be freely chosen and adapted to the specific problem to be simulated. We employ different choices including a Cartesian mesh and a spherical distribution of the mesh-generating points. The implementation presented here represents the first general relativistic moving-mesh code with a dynamical spacetime evolution.

We present the first moving-mesh simulation of a neutron star merger including the late inspiral phase and the post-merger evolution, which we run until roughly 40 milliseconds after merging. The initial mesh setup approximately follows the mass distribution and geometry of the system. For the merger calculation we employ an additional feature of \textsc{Arepo}, namely the adaptive refinement and derefinement of computational cells during the simulation. We find that in the high-density regime criteria to approximately achieve a target mass of the grid cells, which is comparable to the initial setup, work well in merger simulations. Although one cannot define a unique resolution in moving-mesh simulations, the typical cell size in the high-density merger remnant is roughly 150 meters in our simulation. The computational costs for this setup are modest (a few weeks on about 200 cores) and thus even higher resolutions are well achievable. The choice of the initial mesh setup and the refinement/derefinement criteria introduce a certain flexibility, which we will explore in future work with the prospect of further increasing the performance of the tool in merger simulations and possibly identifying choices well-adapted to specific problems or questions in this context.

We analyze the dynamics and the gravitational wave emission of the moving-mesh merger simulation. We find a general agreement with other simulations based on either SPH or static-mesh schemes. In comparison, our moving-mesh calculation seems to preserve the structure of the early post-merger remnant for a longer time. The initial double-core structure persists for more than 20 milliseconds after merging. The quasi-radial oscillation of the remnant is only slowly damped and the profile of the angular velocity evolves on longer time scales. This behavior may prolongate the life time of the remnant although we do not run this specific simulation until the gravitational collapse takes place. Notably, the amplitude of the post-merger gravitational wave emission decreases only slowly and is still large even at the end of the simulation at roughly 40 milliseconds after merging. These characteristics may point to a low numerical viscosity. The frequency of the dominant post-merger oscillation is in good agreement with results from simulations employing other hydrodynamics schemes. At any rate these first results are very encouraging and show that the moving-mesh approach can be very beneficial for the simulation of BNS mergers.

The work presented in this paper will be the basis for more extensive studies and further developments with the new general relativistic moving-mesh code. The inclusion of fully temperature- and composition-dependent EOS tables and neutrino transport is in progress. Other more technical aspects to be addressed in the future concern the Riemann solver, choices for the grid setup, cell merging and splitting, and the atmosphere treatment. The original \textsc{Arepo} code already includes magnetic fields, which may be extended to the relativistic implementation. The currently employed conformal flatness approximation may be replaced by a full solver noting that the infrastructure for communication between the unstructured hydrodynamics grid and overlaid Cartesian grids is already available. The general flexibility and adaptivity of \textsc{Arepo} suggest to employ the relativistic version for other relativistic astrophysical scenarios, e.g. black hole accretion discs and neutron star-black hole mergers.


\section*{Acknowledgements}
We thank the anonymous referee for carefully reading the manuscript and their comments. We thank S.~Blacker, E.~M\"uller, S.~Ohlmann and N.~Stergioulas for helpful discussions. G.L.\ and A.B.\ acknowledge support by the European Research Council (ERC) under the European Union’s Horizon 2020 research and innovation programme under grant agreement No.\ 759253. G.L.\ acknowledges support by the European Research Council (ERC) under the European Union’s Horizon 2020 research and innovation programme (ERC Advanced Grant KILONOVA No. 885281) and by the Deutsche Forschungsgemeinschaft (DFG, German Research Foundation) - MA 4248/3-1. G.L., A.B.\ and F.K.R.\ acknowledge support by Deutsche Forschungsgemeinschaft (DFG, German Research Foundation) - Project-ID 138713538 - SFB 881 (“The Milky Way System”, subproject A10). A.B.\ acknowledges support by DFG - Project-ID 279384907 - SFB 1245 and support by the European
Research Council (ERC) under the European Union’s research and
innovation program (ERC Synergy Grant HEAVYMETAL No. 101071865). A.B.\ and T.S.\ acknowledge support by the State of Hesse within the Cluster Project ELEMENTS. T.S.\ is Fellow of the International Max Planck Research School for Astronomy and Cosmic Physics at the University of Heidelberg (IMPRS-HD) and acknowledges financial support from IMPRS-HD. The work of T.S.\ and F.K.R.\ is supported by the Klaus Tschira Foundation. The simulations of this paper were performed on the Virgo cluster at GSI Helmholtzzentrum f\"ur Schwerionenforschung, Darmstadt.


\section*{Data Availability}

The data underlying this article will be shared on reasonable request to the corresponding author.



\bibliographystyle{mnras}
\bibliography{bibliography}



\appendix

\section{Gravitational wave backreaction formalism}\label{app:GWbackreaction}
In Section \ref{subsec:Backreaction} we briefly discuss how we include GW effects in our code where we use the implementation of~\citet{2007A&A...467..395O}. It follows the analysis presented in \citet{2003PhRvD..68h4001F}, which determines small deviations to the CFC metric. In addition to the CFC equations, we determine a set of potentials ($U_5$, $R$ and $U_7$) by solving the equations
\begin{align}
\Delta U_5 &= -4\pi\sigma, \\
\Delta R   &= -4\pi I^{[3]}_{ij} x^i \partial_j \sigma, \label{DEcfcBR:R}\\
\Delta U_7 &= -4\pi \rho W \sqrt{\gamma} \left( I^{[3]}_{ij} x^i \partial_j U_5 - R \right), \label{DEcfcBR:U7}
\end{align}
where $\sigma=T^{00}+T^{ii}$ and $I^{[3]}_{ij}$ denotes the third (total) time derivative of the quadrupole $I_{ij}$. The quadrupole reads
\begin{equation}
\begin{split}
I_{ij} =& \int \rho W \sqrt{\gamma} \left\{ x^i x^j \left( 1+\frac{w^2}{2}-U_\ast+\epsilon \right) + \frac{11}{21} r^2 w^i w^j \right. \\
& -\frac{4}{7} x^i x^k w^j w^k + \frac{4}{21}w^2 x^i x^j + \frac{11}{21} r^2 x^i \partial_j U_\ast \\
& \left.  - \frac{17}{21} x^i x^j x^k \partial_k U_\ast \right\},
\end{split}
\end{equation}
where $r^2=x^i x^i$, $w^2=w^i w^i$, $w^i=u^i/u^0$ is the coordinate velocity and $U_\ast$ corresponds to the Newtonian gravitational potential \citep{1990MNRAS.242..289B}. Based on the potentials $U_5$, $R$ and $U_7$, the non-conformally flat corrections to the CFC metric read
\begin{align}
h_{00} &= -\frac{4}{5} (1 - 2 U_5) \left( I^{[3]}_{ij} x^i \partial_j U_5 - R \right) -\frac{8}{5} U_7 , \\
h_{ij} &= -\frac{4}{5} I^{[3]}_{ij},
\end{align}
\citep[see][]{2007A&A...467..395O}. The elliptic equations for the potentials  $U_5$, $R$ and $U_7$ are solved with the same multigrid scheme as the CFC equations. The numerical implementation was originally introduced in \citet{2007A&A...467..395O}.

\section{Relativistic shock tube}\label{app:ShockTube}
As an additional test to the TOV evolutions discussed in Sec.~\ref{sec:TOVtests}, we consider a 1D relativistic shock tube, which has been widely employed to test codes \citep[Problem 1 in][]{2003LRR.....6....7M,2015LRCA....1....3M}. The left state ($x\leq0.5$) is described by $p_L=13.33$ and $\rho_L=10$ and the right state ($x>0.5$) by $p_R=10^{-6}$ and $\rho_R=1$. The initial velocity is zero everywhere. The EOS is an ideal gas with an adiabatic index $\Gamma=5/3$. We employ a 1D version of the code and we explicitly adopt a Minkowski metric. We perform both moving- and static-mesh calculations. In all cases, we start with $N$ equally spaced points in a domain $[0,1]$.

\begin{figure}
\includegraphics[width=\columnwidth]{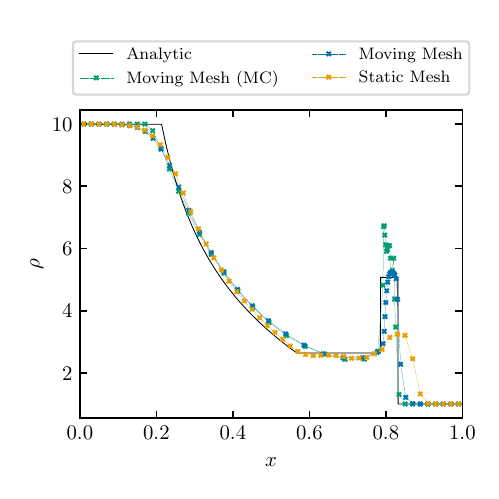}
\caption{Density profile in the relativistic shocktube problem at $t=0.4$. The solid line is the exact solution. The calculations corresponding to the colored lines are listed in the legend. All simulations employ the same initial conditions and resolution. Crosses show the location of mesh cells.}
\label{fig:ShocktubeRho}
\end{figure}

Figure~\ref{fig:ShocktubeRho} shows the density profile of the solution at $t=0.4$ for a low resolution calculation with $N=50$ points. The solid line denotes the exact solution, which we compute with the code provided in the supplemental material of \citet{2003LRR.....6....7M}. The green line corresponds to a moving-mesh calculation with the MC slope limiter. This setup captures the position of the shock front and the contact discontinuity rather accurately but suffers from post-shock oscillations within the dense shell. In fact, we find that some shock tube simulations with moving meshes using MC, as implemented in our code, can even lead to crashes e.g. by forming very small cells in the post-shock region. Post-shock oscillations are known to occur for slope limiters applied in a face-based way \citep[see e.g.][]{Berger2005}. We also report a moving-mesh (blue line) and a static-mesh (orange line) calculation with the standard reconstruction of \textsc{Arepo} (see Eq.~\eqref{ReconOrig}), which replaces the value of the gradient within a cell with a gradient-limited estimate. In Fig.~\ref{fig:ShocktubeRho}, the moving-mesh evolution captures the position and height of the dense shell at $x\approx0.8$ rather well. The static-mesh evolution strongly underestimates the height of this feature, while the dense shell is also smeared out. The static-mesh calculation better resolves the tail of the rarefaction in Fig.~\ref{fig:ShocktubeRho}, probably because it does not lead to a reduction of the resolution by the mesh motion in the rarefaction regime.

To evaluate the convergence properties of the code, we vary $N$ and compute the $L_1$ norm for the density field. For any field $\phi$ (e.g. the rest-mass density $\rho$), we define the $L_1$ norm as
\begin{equation}
    L_1 = \frac{1}{V} \sum_i^N |\phi^\mathrm{num}_i - \phi^\mathrm{exact}_i| V_i,
\end{equation}
where $V=1$ for the interval $[0,1]$, $V_i$ is the size of each cell, $\phi^\mathrm{num}_i$ is the numerical solution for cell $i$ and $\phi^\mathrm{exact}_i$ the exact solution at the center of cell $i$.

\begin{figure}
\includegraphics[width=\columnwidth]{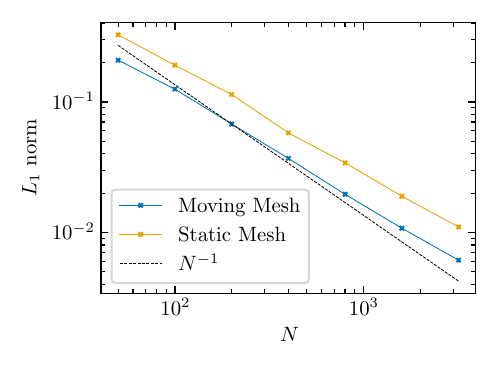}
\caption{$L_1$ norm of the density field at various resolutions for the shocktube problem at $t=0.4$. The blue line corresponds to moving-mesh setups, while the orange line to static-mesh setups. The slope of the dashed line depicts first-order convergence.}
\label{fig:ShocktubeConv}
\end{figure}

Figure~\ref{fig:ShocktubeConv} depicts the $L_1$ norm of the density field as a function of the resolution based on calculations with the standard reconstruction of \textsc{Arepo}. We find that the code exhibits nearly first-order convergence, which is in agreement with the expected convergence in the presence of discontinuities. We note that the $L_1$ error in the density field is consistently lower for the moving-mesh setup compared to the static-mesh evolutions.

\section{Collision of relativistic blast waves}\label{app:BlastWaves}
We consider the collision of two relativistic blast waves \citep[see e.g. Sec.~6.2.3 in][]{2003LRR.....6....7M}. The initial data consist of three states, the left state ($x<=0.1$) where $p_L=1000$, the central state ($0.1<x<=0.9$) where $p_C=0.01$ and the right state ($x>0.9$) where $p_R=100$. Initially, the density is everywhere equal to $1$ and the velocity is zero everywhere. The EOS is an ideal gas with $\Gamma=1.4$, and we adopt outflow boundary conditions, similar to \citet{1996JCoPh.123....1M}. We perform a moving- and a static-mesh simulation, where we initially distribute $800$ equally spaced points in a domain $[0,1]$. Both calculations employ piecewise constant reconstruction to highlight the effect of a moving mesh in a low order scheme. We note that the purpose of this test is to highlight the ability of a moving-mesh approach to capture the density evolution of the system with good accuracy, even when low-order schemes are considered. High order schemes are capable of resolving the structure of the solution \citep[see e.g.][for piecewise parabolic reconstruction and a considerably higher resolution than the one discussed here]{1996JCoPh.123....1M}.

\begin{figure}
\includegraphics[width=\columnwidth]{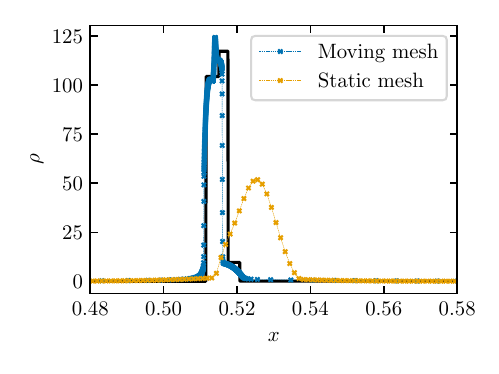}
\caption{Density profile of the blast wave collision problem at $t=0.43$. The solid line is the exact solution. The blue points and line correspond to a moving-mesh calculation, while the orange points and line to a static-mesh calculation with the same initial conditions and resolution. Crosses show the location of mesh cells.}
\label{fig:BlastWaves}
\end{figure}

Figure~\ref{fig:BlastWaves} shows the density profile of the system at $t=0.43$, i.e. after the interaction of the two waves. We show a narrow region which includes the states produced by the collision. The blue points correspond to the moving-mesh calculation, while the orange points display the static-mesh calculation. The solid line is the exact solution based on the calculation in \citet{1996JCoPh.123....1M}. The difference between the moving- and static-mesh calculations is very obvious. The static-mesh setup does not resolve any structure and the states are completely smeared out. Furthermore, the position of the peak is incorrect. Considering the low order of the scheme and the low resolution, this is a reasonable result for the static mesh \citep[see also][]{1996JCoPh.123....1M}. In contrast, the moving-mesh simulation is able to capture the various structures in the collision region relatively accurately. The two distinct states at different heights of the main peak can be identified, while the heights of the states are reproduced reasonably well. Furthermore, the positions of the various states in the collision region are well resolved with only a minor offset. Overall, even though the cells are initially distributed evenly in the simulation domain, the moving mesh follows the fluid motion, which enables it to accurately resolve the very narrow structures which form in this problem.

\section{Numerical setup}\label{app:NumSetup}

\subsection{Binary neutron star merger: resolution study, GW backreaction effect and code performance}\label{subapp:NumSetupRes}

We perform a number of additional simulations for the binary system discussed in Sec.~\ref{sec:NSmergers} to investigate the effect of a number of numerical aspects. These include the hydrodynamical resolution in the simulation (i.e. the number of hydro cells), the resolution employed in the metric solver and the impact of enhancing the strength of the GW backreaction scheme. We consider:
\begin{enumerate}
\item A simulation with a lower hydro resolution. We follow the procedure discussed in Sec.~\ref{subsec:BNSid} to construct the initial hydrodynamical mesh and set $m_\mathrm{cell,0}=2.4\times10^{-6}~M_\odot$. This results in a mesh with roughly $1.2\times10^6$ cells with $\rho>\rho_\mathrm{thr}$, which corresponds to about $\approx70\%$ of the mass resolution of the original simulation. Other choices of the numerical setup (e.g.\ the metric grid resolution) remain identical to the original simulation.
\item A simulation with an hydrodynamical setup identical to the one described in Sec.~\ref{subsec:BNSid}, but the metric field equations are solved on an overlaid uniform Cartesian grid with $193^3$ points and a cell size of $0.533~M_\odot$ (i.e. the metric grid resolution is higher than in the original simulation).
\item An identical setup to the one discussed in Sec.~\ref{subsec:BNSid}, where we enhance the strength of the GW backreaction scheme. This is achieved by multiplying the derivatives of the quadrupole moment entering the source terms in equations \eqref{DEcfcBR:R}-\eqref{DEcfcBR:U7} by a factor $2$ (see Appendix \ref{app:GWbackreaction}). Hence, we effectively account for the observation that the quadrupole formula may underestimate the GW signal amplitude by some $10\%$ \citep[see e.g.][]{2005PhRvD..71h4021S,2022EPJA...58...74D}. A factor $2$ is chosen to safely bracket the potential impact.
\end{enumerate}

\begin{figure}
\includegraphics[width=\columnwidth]{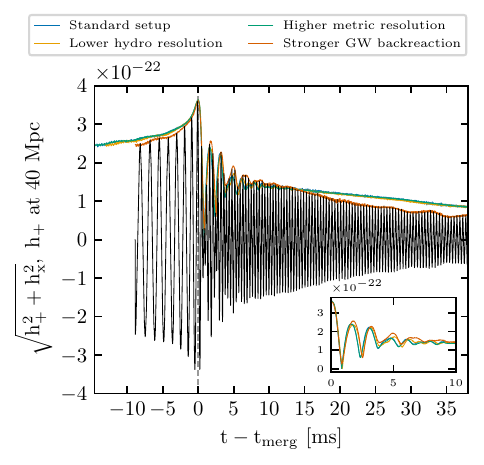}
\caption{Gravitational wave amplitude $\sqrt{h_\mathrm{+}^2+h_\mathrm{x}^2}$ versus time for the simulations outlined in the legend. For the simulation with an enhanced backreaction, the plus polarization is also shown (see main text). A smaller panel which depicts a zoomed in version of the first $10$~ms in the post-merger phase is also included. Note that $h_\mathrm{+}$ is shown, unlike Fig.~\ref{fig:BNSGWsignal} which displays $h_\mathrm{+}^\mathrm{1.4}$.}
\label{fig:BNS_GWamp_res}
\end{figure}

Figure~\ref{fig:BNS_GWamp_res} shows the GW strain amplitude for the simulation discussed in Sec.~\ref{sec:NSmergers}, as well as all three additional simulations presented here. We present the amplitude $\sqrt{h_\mathrm{+}^2+h_\mathrm{x}^2}$ for all the simulations, while for case (iii) we also include the plus polarization of the GW signal, $h_\mathrm{+}$. We emphasize that for case (iii) we extract $h_\mathrm{+}$ from the quadrupole moment without including the enhancement factor $2$. This factor is only included in the quadrupole moment derivative terms in equations \eqref{DEcfcBR:R}-\eqref{DEcfcBR:U7}. As a result, all the simulations reported in Fig.~\ref{fig:BNS_GWamp_res} have comparable amplitudes. The subpanel within Fig.~\ref{fig:BNS_GWamp_res} zooms in on the first $10$~ms of the post-merger phase. Like this the characteristics of the GW amplitudes are easier to read off. In all cases, the signals are shifted such that $t-t_\mathrm{merg}=0$ corresponds to the moment of merging in the respective simulation.

Comparing the simulations which have a different hydrodynamical or metric resolution to the standard setup (i.e.\ cases (i) and (ii)), we find a stronger dependence on the hydrodynamical resolution, but overall
good agreement with the original simulation. Reducing the hydrodynamical resolution, the stars merge $\approx1.2$~ms earlier compared to the standard simulation. Increasing the metric resolution has a milder effect and the stars merge $\approx0.3$~ms later than the original setup\footnote{As said, we align the simulations to merge at $t-t_\mathrm{merg}=0$, which is why these
differences cannot be easily read off from the plot. But note that the various lines start at different times in Fig.~\ref{fig:BNS_GWamp_res}.}. Similarly, the GW signal in the early post-merger phase remains practically unaffected by the increase in the metric resolution and only a very mild effect can be seen in the simulation with lower hydrodynamical resolution. On longer timescales of several ten milliseconds, the GW signal still looks very similar in all three simulations (comparing the standard setup, case (i) and case (ii)) and the damping timescale is practically the same. The double-core structure persists for $\approx1.5$~ms less in the run with the lower hydro resolution compared to the standard setup, while the higher metric resolution has practically no effect on when the cores merge.

Enhancing the strength of the GW backreaction scheme (case (iii)) has a more noticeable effect on the evolution, which is  expected since GWs are an important damping mechanism. The inspiral time is reduced by $\approx5.4$~ms. The impact of a stronger GW backreaction in the early post-merger phase is a bit more pronounced than the effects considered in cases (i) and (ii). We note however that we changed the hydro resolution only within a relatively small range and thus we cannot exclude that the choice of the hydrodynamical resolution may actually have a larger effect. Furthermore, the damping timescale of the GW emission is shorter than our standard setup, which becomes clearly visible at late times in Fig.~\ref{fig:BNS_GWamp_res}. We emphasize that the damping of the post-merger signal remains very slow and we extract a damping timescale of $\tau_\mathrm{peak}\approx31$~ms based on the model in \citet{2022PhRvD.105d3020S}. In addition, we find that the two cores in the postmerger phase merge after roughly $12$~ms. Hence, in the set of neutron star merger simulations discussed in this work, certain dynamical features in the post-merger phase seem to consistently persist for longer timescales compared to simulations with other numerical schemes.

Finally, we comment on the effect of the different resolutions in the hydro scheme (case(i)) and metric solver (case (ii)) on the  computational costs. \textsc{Arepo} is written in C and parallelized for distributed memory systems with MPI. The metric solver is written in Fortran and parallelized with OpenMP. We emphasize that the numbers quoted here are only representative of specific simulations presented here. The distribution of mesh-generating points can affect how much time is required for the various operations within a timestep, e.g.\ for the mesh construction. As measure of the computational costs, we consider the wall clock time required for a timestep where the metric field equations are explicitly solved and a timestep where the metric fields are only extrapolated (see Sec.~\ref{subsec:TimeUpdate}). The second value is representative of how much time is required for the various operations related to the hydrodynamical evolution and the mesh construction\footnote{Extrapolating the metric is very cheap such that it hardly affects the numbers.}, while the difference between the two numbers captures the costs for metric-related operations (namely the metric solver and the treewalk to place metric grid points in the Voronoi hydrodynamical mesh). We extract these times as an average of the first $100,000$ steps in each simulation to get representative values. We exclude timesteps where a snapshot of the full 3D simulation data is produced, which is a major I/O operation. For the standard setup, we find that the average timestep where the metric field equations are explicitly solved takes $\approx8.1$~s, while extrapolating the metric fields reduces the time to $\approx5.9$~s. For the simulation with lower hydro resolution, these numbers are $\approx5.7$~s and $\approx3.4$~s, respectively. The metric resolution is the same in both simulations, and thus the time spend for the metric solution is roughly constant (about $2.2$~s). We note that our standard setup employs $192$ cores, while the simulation with a lower hydrodynamical simulation uses $184$ cores. Finally, increasing the metric grid from $129^3$ (standard setup) to $193^3$ (case (ii)) raises the time spent on metric-related operations from $\approx2.2$~s to $8.2$~s. Hence, considering only the time spent in the metric solver, we find that in the run with a better metric resolution it increases by a factor of $\approx3.8$, which indicates a very good scaling for the metric solver, since a perfect scaling would be $193^3/129^3\approx3.35$.

\subsection{Frequency of solving the metric field equations}\label{subapp:NumSetupFreq}
\begin{figure}
     \begin{subfigure}[b]{\columnwidth}
         \centering
            \includegraphics[width=\textwidth]{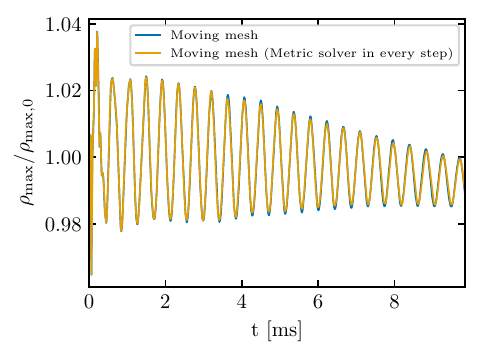}
            \caption{}\label{fig:MetricFreqA}
    \end{subfigure}

     \begin{subfigure}[b]{\columnwidth}
         \centering
            \includegraphics[width=\textwidth]{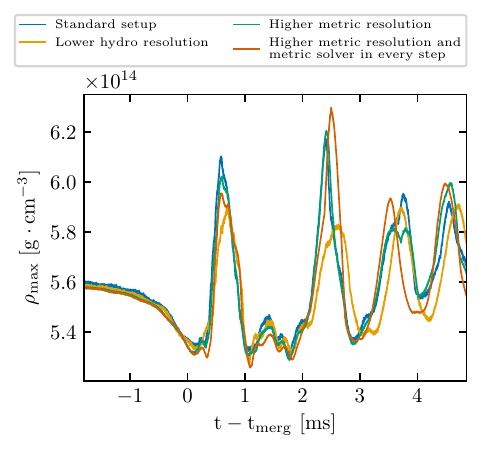}
            \caption{}\label{fig:MetricFreqB}
    \end{subfigure}
\caption{Impact of solving the metric field equations in every substep of the time integration scheme on the density evolution. Panel (a): Moving-mesh simulations of TOV stars. The blue line corresponds to the low resolution moving-mesh setup discussed in Sec.~\ref{subsec:TOVGRid}, while the orange line is the same setup where the metric fields are explicitly computed in every substep. Panel (b): Effect on BNS simulations. The original setup from Sec.~\ref{sec:NSmergers} is shown in blue together with three additional calculations with lower hydro resolution (orange), higher metric resolution (green) and higher metric resolution combined with solving the field equations in every substep (red) .}
\label{fig:MetricFreq}
\end{figure}

In Section \ref{subsec:TimeUpdate}, we discuss how frequently we solve the metric field equations within the time integration scheme. Here, we investigate the impact of this choice to justify the approach. Focusing on the evolution of TOV stars, we perform a moving-mesh simulation with the low-resolution setup described in Sec.~\ref{subsec:TOVGRid}, where we solve the metric field equations in every substep of the time integration. We present the maximum density evolution from this simulation in Fig.~\ref{fig:MetricFreqA} (orange line). The blue line displays the density evolution for the original setup where we solve the metric field equations according to the description in Sec.~\ref{subsec:TimeUpdate}, i.e.\ employing an extrapolation for a subset of timesteps. We find that the frequency of solving the metric field equations has practically no impact on the evolution. Both the amplitude of the oscillation, as well as the dominant frequency, remain practically unaffected by solving the metric field equations more frequently.

In addition, we evolve the standard resolution moving-mesh setup from Sec.~\ref{subsec:TOVGRid}, but we explicitly solve the field equations in the first substep of every Runge--Kutta timestep, i.e.~we call the metric solver more frequently compared to our default settings. We again find that it does not affect the overall evolution and the differences are even less visible than in Fig.~\ref{fig:MetricFreqA}.

Furthermore, we evaluate the effect of explicitly solving the metric field equations in every substep of the time integration scheme in the case of a BNS merger. We consider the setup described in case (ii) in Appendix~\ref{subapp:NumSetupRes} (i.e. the metric grid resolution is $0.533~M_\odot$) and simulate the interval from $1.75$~ms before the moment of merging up to $4.85$~ms after the time of merging. The metric fields change rapidly during merging. Hence, this is arguably the most challenging interval during the simulation to accurately estimate the values of the metric fields based on extrapolation. Figure~\ref{fig:MetricFreqB} shows the evolution of the maximum density close to the moment of merging for four different simulations. In addition, we overplot three more simulations: the original BNS simulation discussed in Sec.~\ref{sec:NSmergers} (with a lower metric resolution), as well as the setups with lower hydro resolution (case (i)) and higher metric resolution (case (ii)) from Appendix~\ref{subapp:NumSetupRes}.

The set of simulations presented in Fig.~\ref{fig:MetricFreqB} allows to judge the impact of solving the metric field equations more frequently in comparison to other numerical choices like a different hydro or metric resolution. We find that the two calculations with a metric resolution of $0.533~M_\odot$ progress rather similarly, but solving the metric field equations more frequently does have a recognizable impact on the maximum density evolution. This result suggests that BNS simulations can benefit up to some extent from solving the metric field equations more frequently, at least during the merger stage. However, we note that the differences are smaller than those that one obtains from a change of the hydro resolution by $30\%$ (considering the number of cells resolving stellar matter). This justifies a practical approach where saving computational resources by calling the metric solver less often can be used for a better resolution.

\bsp	
\label{lastpage}
\end{document}